\documentclass[journal,twoside]{IEEEtran}
\usepackage{cite}
\usepackage{amsmath,amssymb,amsfonts}
\usepackage{algorithmic}
\usepackage{graphicx}
\usepackage{subfigure}
\usepackage{textcomp}
\usepackage{color}
\def\BibTeX{{\rm B\kern-.05em{\sc i\kern-.025em b}\kern-.08em
    T\kern-.1667em\lower.7ex\hbox{E}\kern-.125emX}}
\begin{document}
\title{Performance Considerations of Thin Ferroelectrics ($\sim$ 10 nm HfO$_{2}$, $\sim$ 20 nm PZT) FDSOI NCFETs for Digital Circuits at Reduced Power Consumption}
\author{Shruti Mehrotra, \IEEEmembership{Student Member, IEEE}, and S. Qureshi, \IEEEmembership{Senior Member, IEEE}
\thanks{Manuscript originally submitted on March 7, 2018 and resubmitted on July 2, 2018.}
\thanks{The authors are with the Department of Electrical Engineering, Indian Institute of Technology, Kanpur, India (e-mail: mshruti@iitk.ac.in, qureshi@iitk.ac.in).}
}

\maketitle

\begin{abstract}
The paper presents simulation study of thin ferroelectrics (Si doped HfO$_{2}$, PZT) PGP FDSOI NCFETs at circuit level for high performance, low V$_{DD}$ low-power digital circuits. The baseline PGP FDSOI MOSFET has 20 nm metal gate length with supply voltage varying from 0.5 V to 0.9 V. The circuits studied were 3-stage CMOS ring oscillator, NAND-2 and NOR-2 gates at a frequency of 20 GHz. The paper shows that HfO$_{2}$ FDSOI NCFET based NAND-2 gates can provide significant reduction in average power consumption, which was $\sim$66\% that of baseline FDSOI MOSFET based NAND-2 gates for comparable performance. For the same performance, the average power consumption for PZT FDSOI NCFET based NAND-2 gate was $\sim$86\% that of baseline FDSOI MOSFET based NAND-2 gate. The power-delay product of HfO$_{2}$ FDSOI NCFET based gates was found to be $\sim$24\% lower than baseline FDSOI MOSFET based gates and that of PZT FDSOI NCFET based gates was found to be $\sim$21\% less than that of baseline FDSOI MOSFET based gates. The performance of HfO$_{2}$ FDSOI NCFET based gates with increased fan-in and fan-out was also found to be superior to PZT FDSOI NCFET based gates and baseline FDSOI MOSFET based gates.
\end{abstract}


\section{Introduction}
\label{sec:introduction}
\IEEEPARstart{A}{s} the technology continues to scale down, the use of ferroelectrics in MOSFETs to provide negative capacitance (NC) effect holds significance in achieving sub-60 mV/decade subthreshold swing (SS) \cite{Sayeef_ACSNano_2008}. The benefits of using NCFETs for low-power operation have been highlighted in several studies for different ferroelectrics, viz., PZT \cite{Alam_PZT,Datta_PZT,Majumdar_TED_2016,Saeidi_TED_2016,Li_delay}, BTO \cite{Alam_PZT} and doped HfO$_{2}$ (Si, Zr, Y, Gd, La) \cite{Esseni_EDL,Ota_JJAP_2016,AIKhan_APL_2011,Sayeef_FDSOI}. The ferroelectrics of greatest interest are PZT and doped HfO$_{2}$ (Si, Zr). The use of doped HfO$_{2}$ as a ferroelectric in NCFETs is attractive not only because it is compatibile with existing fabrication techniques, but also for its high switching speed \cite{Sayeef_speed}. Henceforth, Si doped HfO$_{2}$ used as ferroelectric in this work is referred to as HfO$_{2}$. Recent studies have also reported fabrication of NCFETs using standard gate-last process \cite{Gate_last,PZT_Gate-last}. Although several studies have been reported on these ferroelectrics, to the best of our knowledge, a comprehensive study of HfO$_{2}$ FDSOI NCFETs and PZT FDSOI NCFETs at the gate level for 20 nm gate length has not been reported. In this paper, a detailed study of performance and average power consumption of FDSOI NCFETs with a thin layer of HfO$_{2}$ or PZT in the gate stack is reported for logic circuits. The baseline FDSOI MOSFET based gates are operated at low V$_{DD}$ to save power. But, this causes performance degradation. The solution to this problem is using FDSOI NCFETs at low V$_{DD}$ for logic gates which achieve high performance at low average power consumption. The FDSOI NCFETs have been used to build 3-stage CMOS ring oscillators and 2-input universal gates, namely, NAND and NOR. A comparison has been drawn with the performance of the same circuits made using baseline FDSOI MOSFETs. Further, Power-Delay Product (PDP) of FDSOI NCFET based gates has been evaluated and compared with that of FDSOI MOSFET based gates. The damping effect of the ferroelectrics is ignored in this work for prediction of best device performances \cite{Damping_const},\cite{NDR_FinFET_TED2017}. 

This paper is organized as follows. In Section II the device details are given, Section III covers evaluation of FDSOI NCFET based logic gates, Section IV compares the performance of HfO$_{2}$ FDSOI NCFET based gates with PZT FDSOI NCFET based gates followed by conclusion in Section V.
\section{Device Details}
\begin{figure}[!ht]
\setlength{\abovecaptionskip}{-3.5pt}
\vspace{-0.1in}
     \centering
     \subfigure[]{\includegraphics[trim=0cm 0.5cm 0cm 0cm clip=true,width=4.0cm,height=3.85cm]{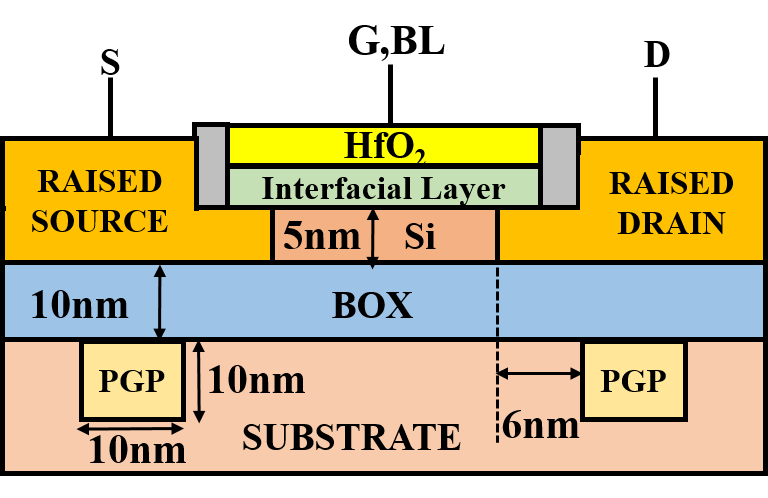}\label{str1}}
		 \subfigure[]{\includegraphics[trim=0cm 3.55cm 0cm 0cm clip=true,width=4.5cm,height=4.4cm]{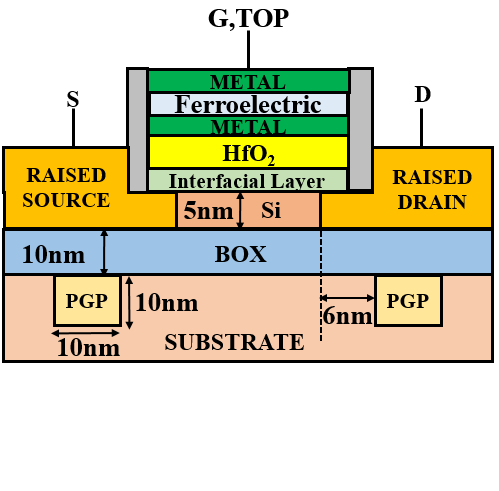}\label{str_PGP}}
     \caption{(a) Schematic of PGP FDSOI MOSFET (baseline MOSFET). The metal gate length is 20 nm. HKMG is used as gate stack with an EOT of 0.9 nm. (b) Schematic of PGP FDSOI NCFET with ferroelectric in gate stack. For baseline FDSOI MOSFET, V$_{GS,TOP}$=V$_{GS,BL}$.}
     \label{str}
\end{figure}

FDSOI MOSFETs with Partial Ground Planes (PGPs) were used as baseline devices as PGPs are known to improve Drain Induced Barrier Lowering (DIBL) behaviour as explained in subsequent subsection B. The structure of baseline FDSOI MOSFET is shown in Fig. \ref{str1}. The devices had a metal gate length of 20 nm and a HKMG gate stack with an EOT of 0.9 nm. The silicon layer that forms the channel was 5 nm thick and was intrinsically doped ($\sim$10$^{15}$ cm$^{-3}$). The BOX was 10 nm thick. The source and drain regions were degenerately doped with a doping concentration of $\sim$10$^{20}$ cm$^{-3}$. The PGPs were heavily doped ($\sim$10$^{20}$ cm$^{-3}$) regions located 6 nm from the source/channel and drain/channel junctions and had the same dimensions as in \cite{SYanagi_EDL_2001}. FDSOI NCFETs studied in this paper had a Metal-Ferroelectric-Metal-Insulator-Semiconductor (MFMIS) structure, as shown in Fig. \ref{str_PGP}. The internal metal layer in the MFMIS structure helps in providing a uniform electric field to the underlying baseline device, ignoring charge trapping or detrapping \cite{Rusu_IEDM_2010},\cite{Khan_IEDM_2011}.  

\begin{figure}[!ht]
\setlength{\abovecaptionskip}{-3.5pt}
\vspace{-0.1in}
     \centering
     \subfigure[]{\includegraphics[trim=2cm 1.8cm 2.8cm 2cm clip=true,width=4.25cm,height=4.2cm]{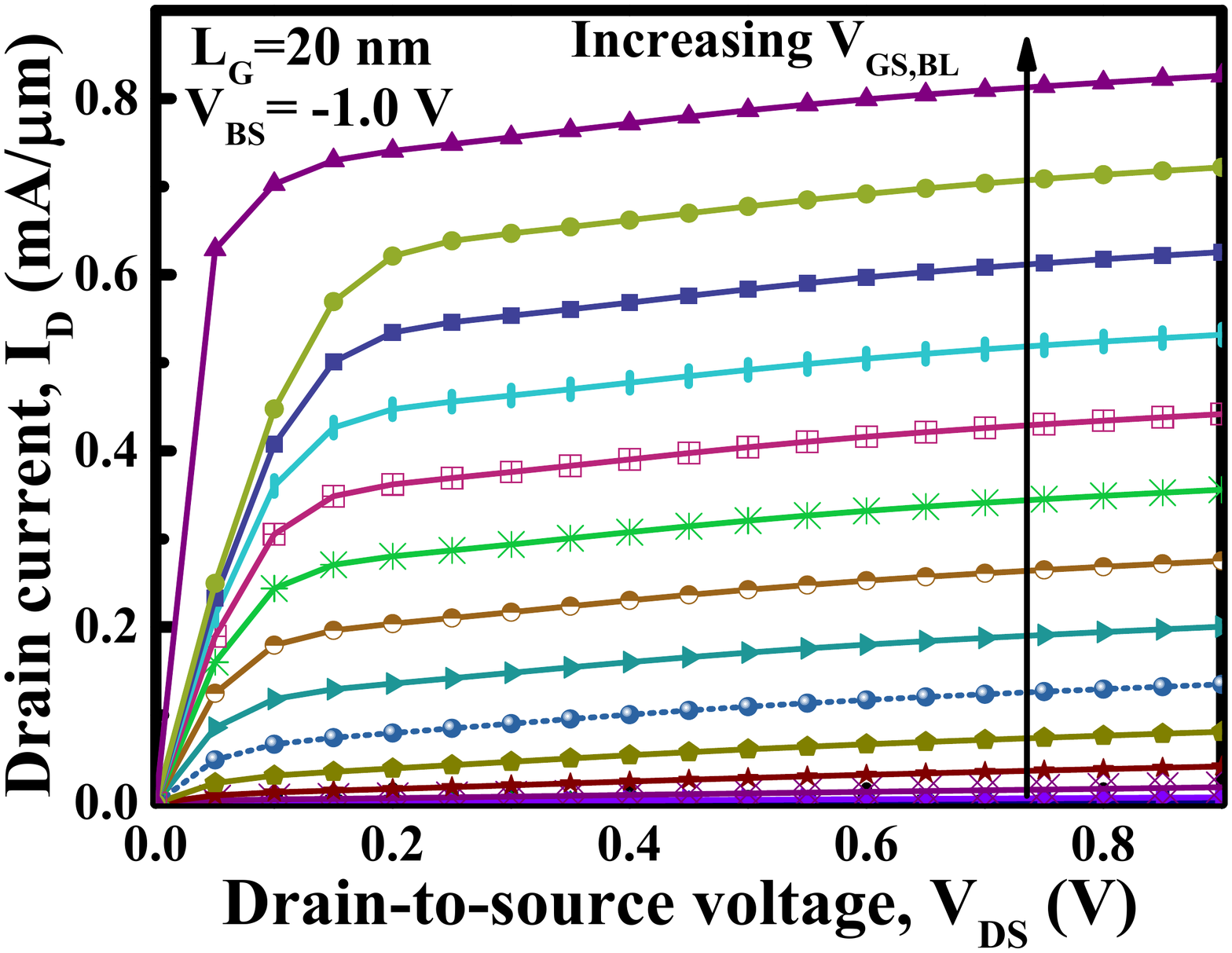}\label{n-ivd_bl}}
		 \subfigure[]{\includegraphics[trim=2cm 1.8cm 2.8cm 2cm clip=true,width=4.25cm,height=4.2cm]{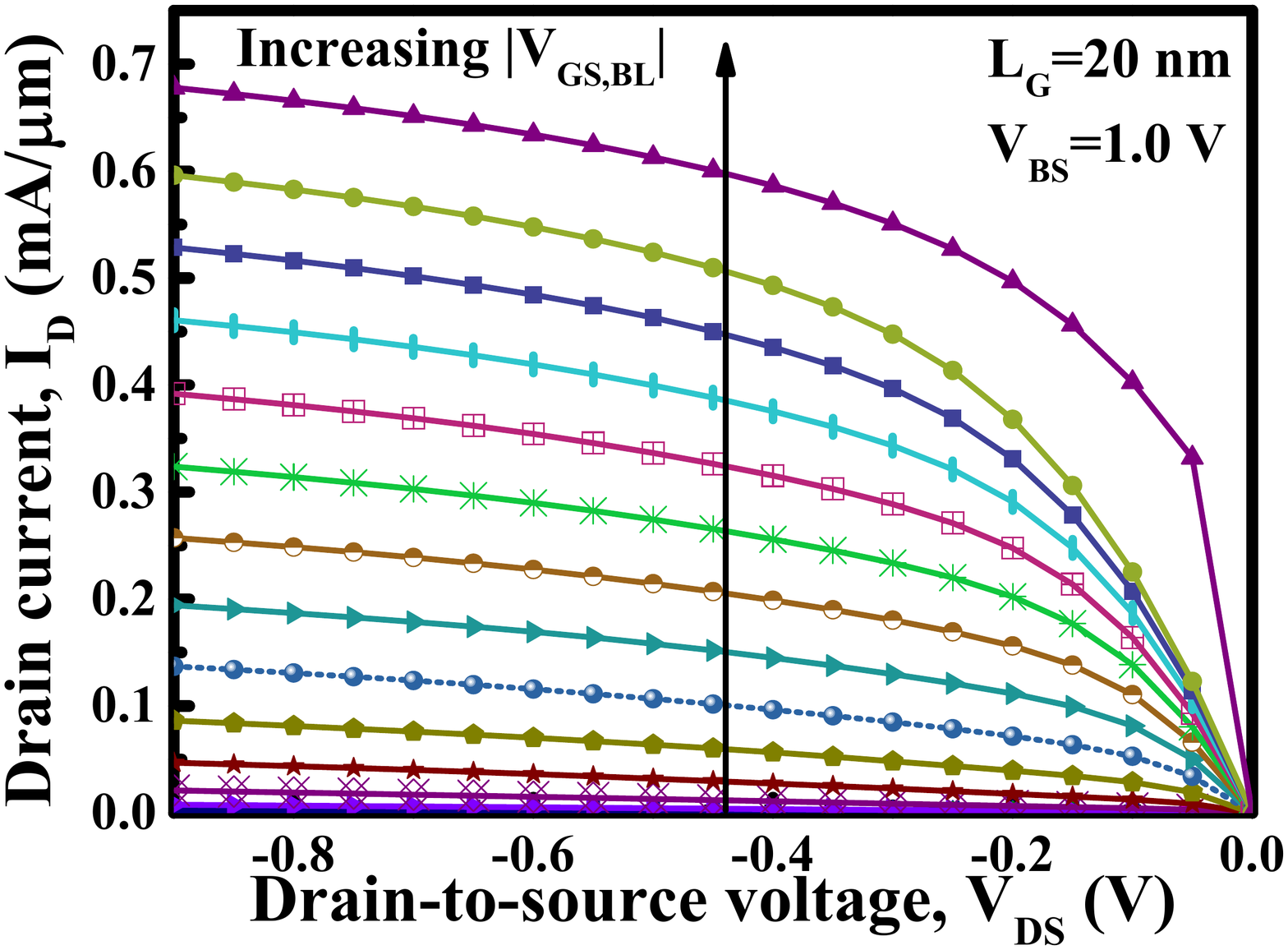}\label{p-ivd_bl}}
     \caption{I$_{D}$-V$_{DS}$ characteristics of baseline (a) FDSOI n-MOSFETs and, (b) FDSOI p-MOSFETs, with increasing V$_{GS,BL}$. $\mid$$V_{BS}$$\mid$=1 V.}
     \label{ivd_bl}
\end{figure}

\begin{figure}[!ht]
\setlength{\abovecaptionskip}{-3.5pt}
     \centering
     \subfigure[]{\includegraphics[trim=2cm 1.6cm 4cm 2cm clip=true, width=4.25cm,height=4.0cm]{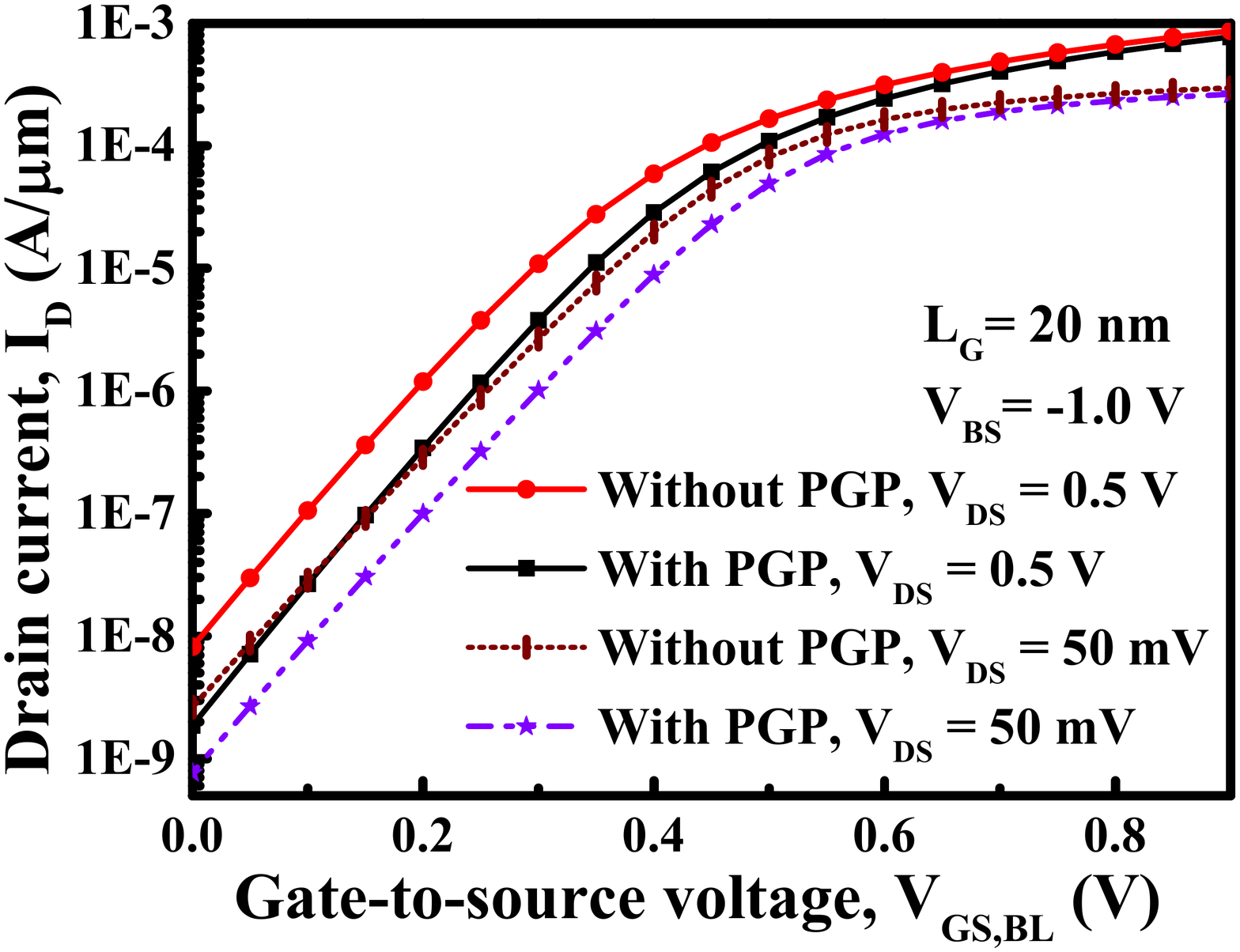}\label{n-DIBL}}
		 \subfigure[]{\includegraphics[trim=1cm 1.6cm 4cm 2cm clip=true, width=4.25cm,height=4.0cm]{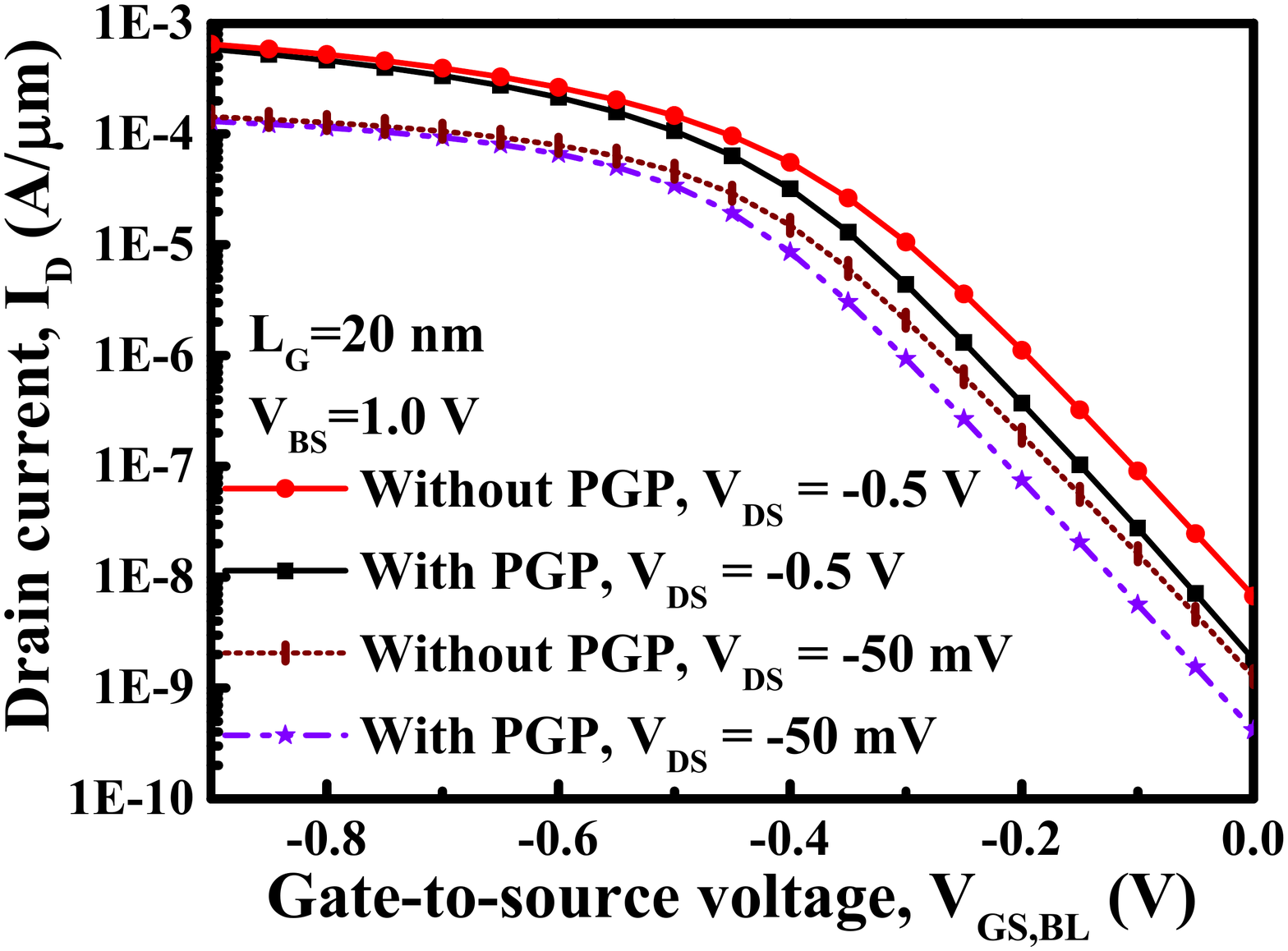}\label{p-DIBL}}
     \caption{Reduction of DIBL in (a) FDSOI n-MOSFETs (from 178 mV/V to 156 mV/V), and (b) FDSOI p-MOSFETs (from 333 mV/V to 288 mV/V), with PGPs. $\mid$V$_{BS}$$\mid$ = 1 V. }
\end{figure}

\subsection{Baseline device}
To realize the baseline device, a structure as proposed in \cite{Liu_2013_IEDM} was first simulated at 20 nm gate length in Silvaco ATLAS TCAD \cite{Silvaco_ATLAS}. The mobility models used were Lombardi mobility model and high field mobility model. Fermi-Dirac statistics, Auger and Shockley-Read-Hall recombination models were also invoked. Since the silicon thickness was 6 nm, quantum confinement effect was also considered. The BOX thickness was 25 nm as proposed in \cite{Liu_2013_IEDM}. After calibration with \cite{Liu_2013_IEDM}, the thicknesses of the silicon layer and BOX were reduced to 5 nm and 10 nm respectively and PGPs were introduced as per \cite{SYanagi_EDL_2001}. A constant reverse bias of $\mid$1 V$\mid$ was applied to achieve better front gate control \cite{Majumdar_TED_2009}. The threshold voltage was determined based on the constant current method, at a drain current of 100 $\mu$A/$\mu$m. The threshold voltage of baseline FDSOI n-MOSFET was 0.5 V and that of baseline FDSOI p-MOSFET was -0.5 V. The output characteristics of the baseline FDSOI n-MOSFET and FDSOI p-MOSFET are shown in Fig. \ref{ivd_bl}.

\begin{figure}[!ht]
\setlength{\abovecaptionskip}{-3.5pt}
\vspace{-0.18in}
     \centering
     \subfigure[]{\includegraphics[trim=2cm 1.8cm 2.8cm 0.5cm clip=true,width=4.25cm,height=4.2cm]{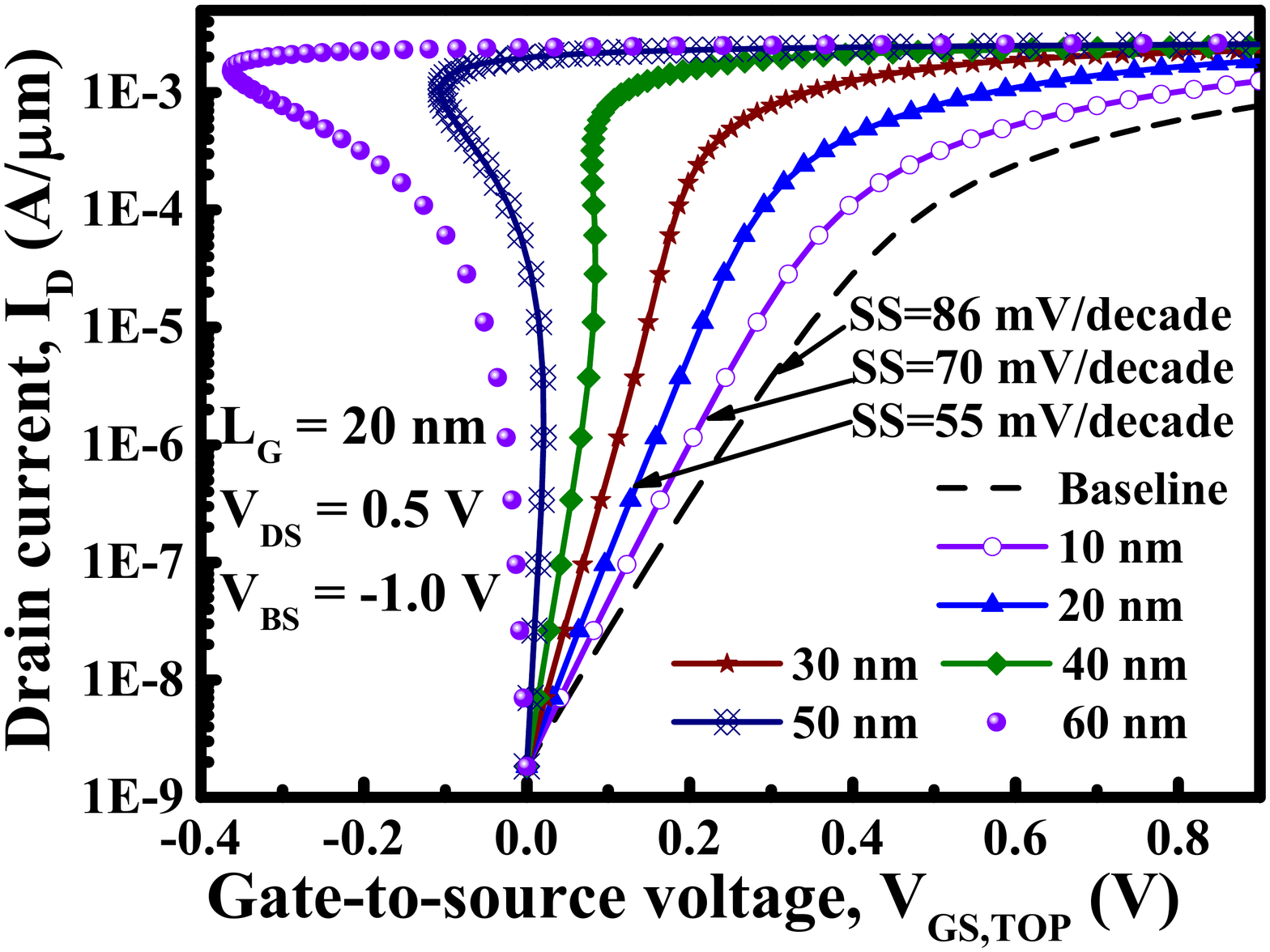}\label{2a}}
		 \subfigure[]{\includegraphics[trim=2cm 1.8cm 3.8cm 0.5cm clip=true,width=4.25cm,height=4.2cm]{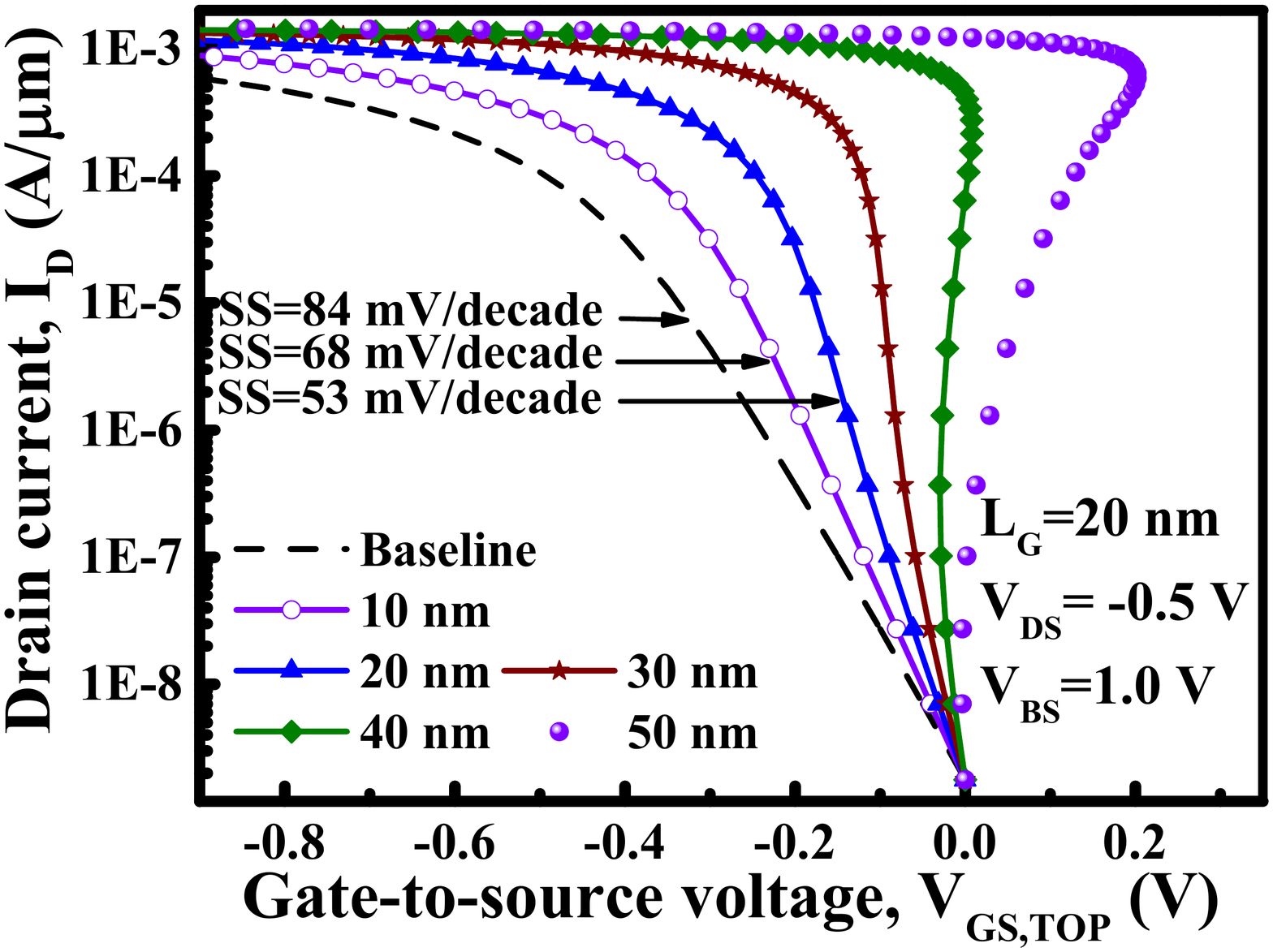}\label{2b}}
     \caption{I$_{D}$-V$_{GS}$ characteristics of FDSOI (a) n-NCFETs and, (b) p-NCFETs, at $\mid$$V_{DS}$$\mid$ of 0.5 V, with varying T$_{FE}$ of HfO$_{2}$. The subthreshold swing (SS) improvement can be clearly seen in FDSOI NCFETs. FDSOI n-NCFET with T$_{FE}$=10 nm has SS=70 mV/decade. FDSOI p-NCFET with T$_{FE}$=10 nm has SS=68 mV/decade. $\mid$$V_{BS}$$\mid$=1 V.}
     \label{ivg_hfo2}
\end{figure}

\begin{figure}[!ht]
\setlength{\abovecaptionskip}{-3.5pt}
\vspace{-0.18in}
     \centering
     \subfigure[]{\includegraphics[trim=2cm 1.8cm 4cm 2cm clip=true, width=4.25cm,height=4.0cm]{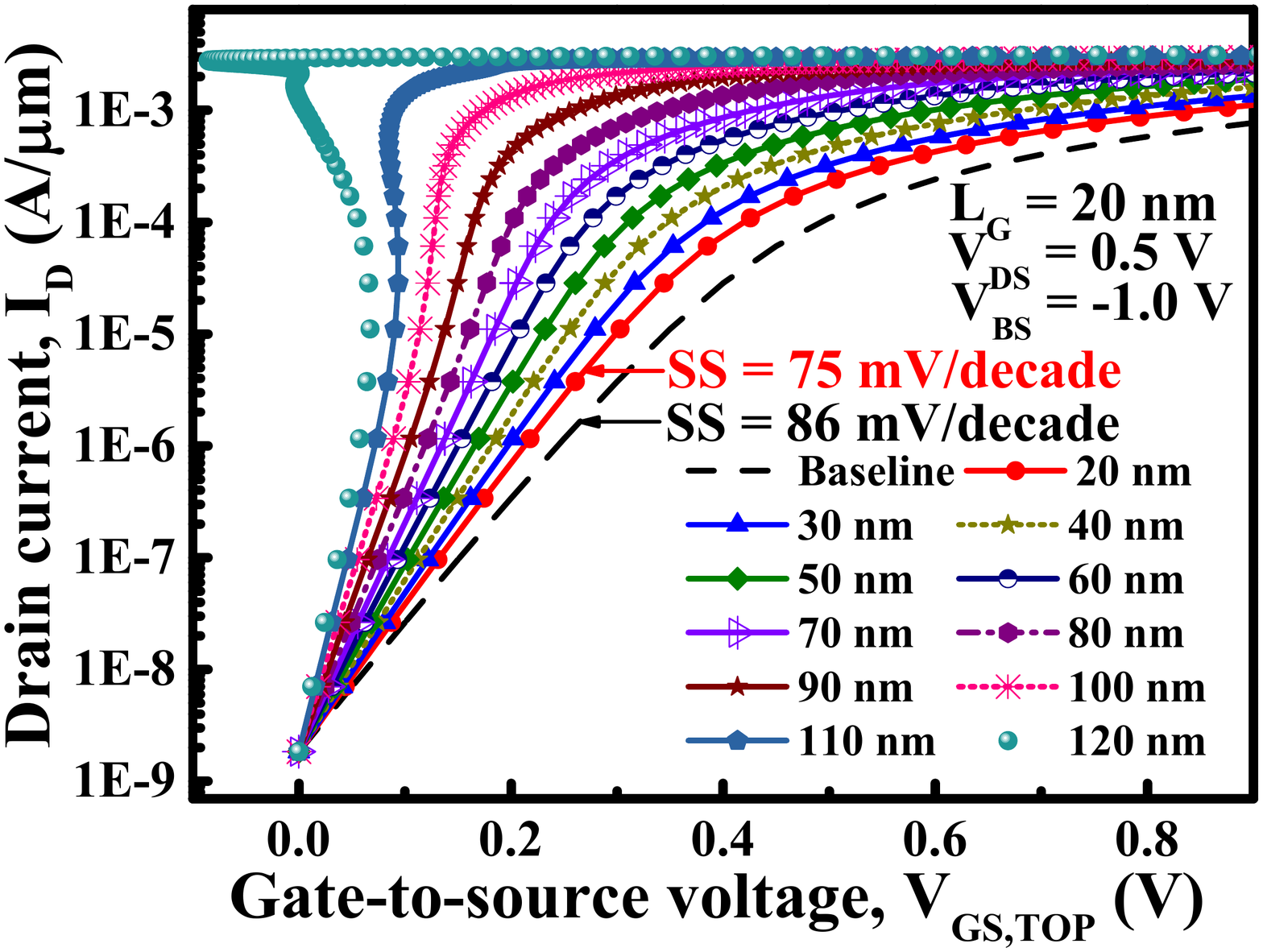}\label{6a}}
		 \subfigure[]{\includegraphics[trim=1cm 1.8cm 4cm 2cm clip=true, width=4.25cm,height=4.0cm]{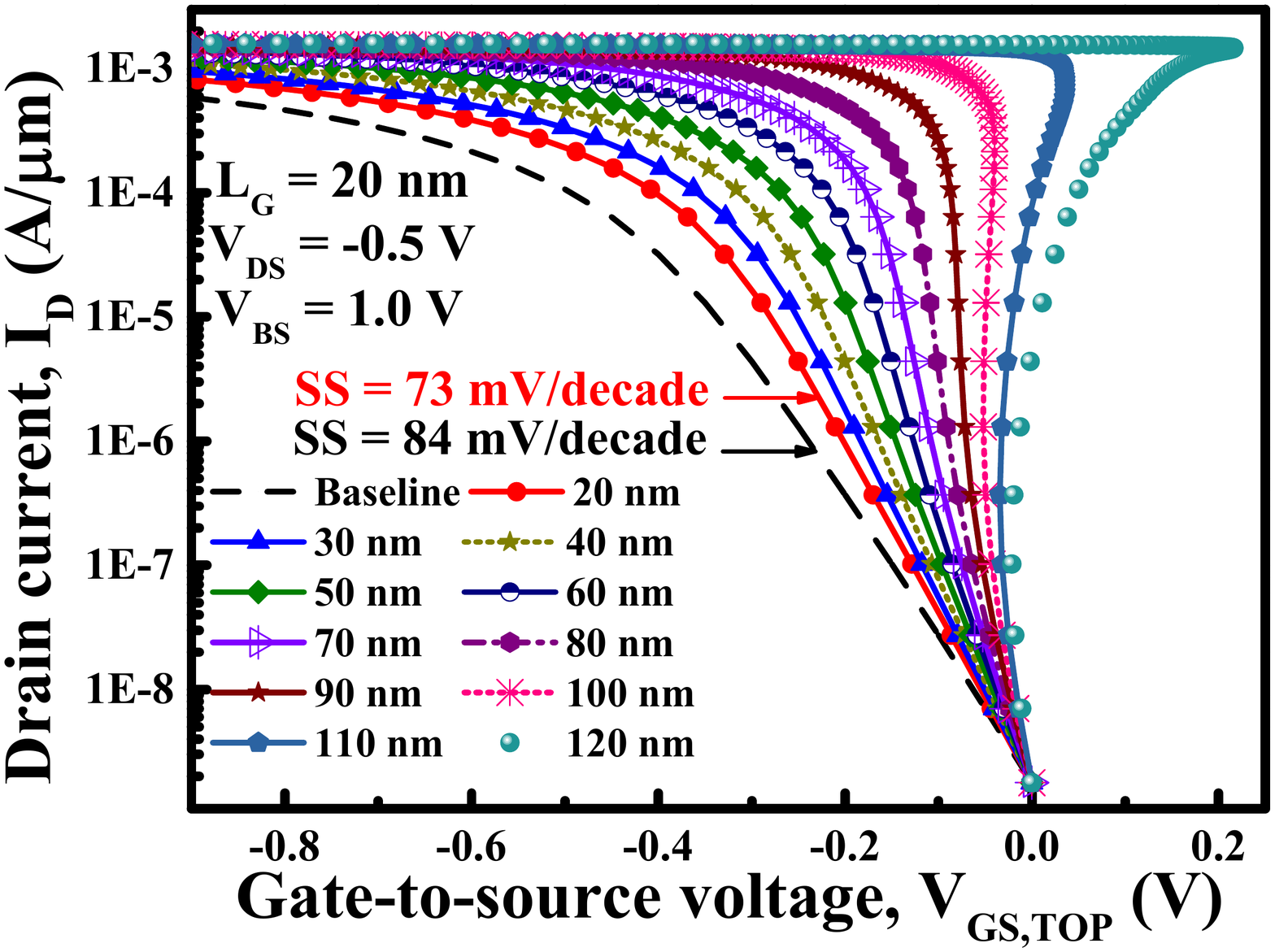}\label{6b}}
     \caption{I$_{D}$-V$_{GS}$ characteristics of FDSOI (a) n-NCFETs and, (b) p-NCFETs, at $\mid$$V_{DS}$$\mid$ of 0.5 V, with varying T$_{FE}$ of PZT. The SS improvement can be clearly seen in FDSOI NCFETs. FDSOI n-NCFET with T$_{FE}$=20 nm has SS=75 mV/decade. FDSOI p-NCFET with T$_{FE}$=20 nm has SS=73 mV/decade. $\mid$$V_{BS}$$\mid$=1 V.}
     \label{ivg_pzt}
\end{figure}

\subsection{Role of PGPs}
PGPs have been reported to cause a reduction in DIBL as they help in keeping the gate-induced field high in the silicon layer of an SOI MOSFET \cite{SYanagi_EDL_2001,Sajad_TED_2010,2017S3S}. Fig. \ref{n-DIBL} shows $\sim$12.35\% reduction in DIBL (from 178 mV/V to 156 mV/V) achieved after incorporation of PGPs in baseline FDSOI n-MOSFETs. Fig. \ref{p-DIBL} shows $\sim$13.5\% reduction in DIBL from 333 mV/V to 288 mV/V, due to PGPs in baseline FDSOI p-MOSFETs. The effect of PGPs in FDSOI n-NCFETs was also analysed and it was found that the reduction in DIBL in FDSOI n-NCFETs was $\sim$33\% (from 35.3 mV/V to 23.5 mV/V). The PGPs can be self-aligned with the gate as described in \cite{Colinge,GP_Jaggu,Sajad_TED_2010}. 

\begin{figure}[!ht]
\setlength{\abovecaptionskip}{-3.5pt}
\vspace{-0.15in}
     \centering
     \subfigure[]{\includegraphics[trim=2cm 1.8cm 4cm 2cm clip=true, width=4.25cm, height=4.0cm]{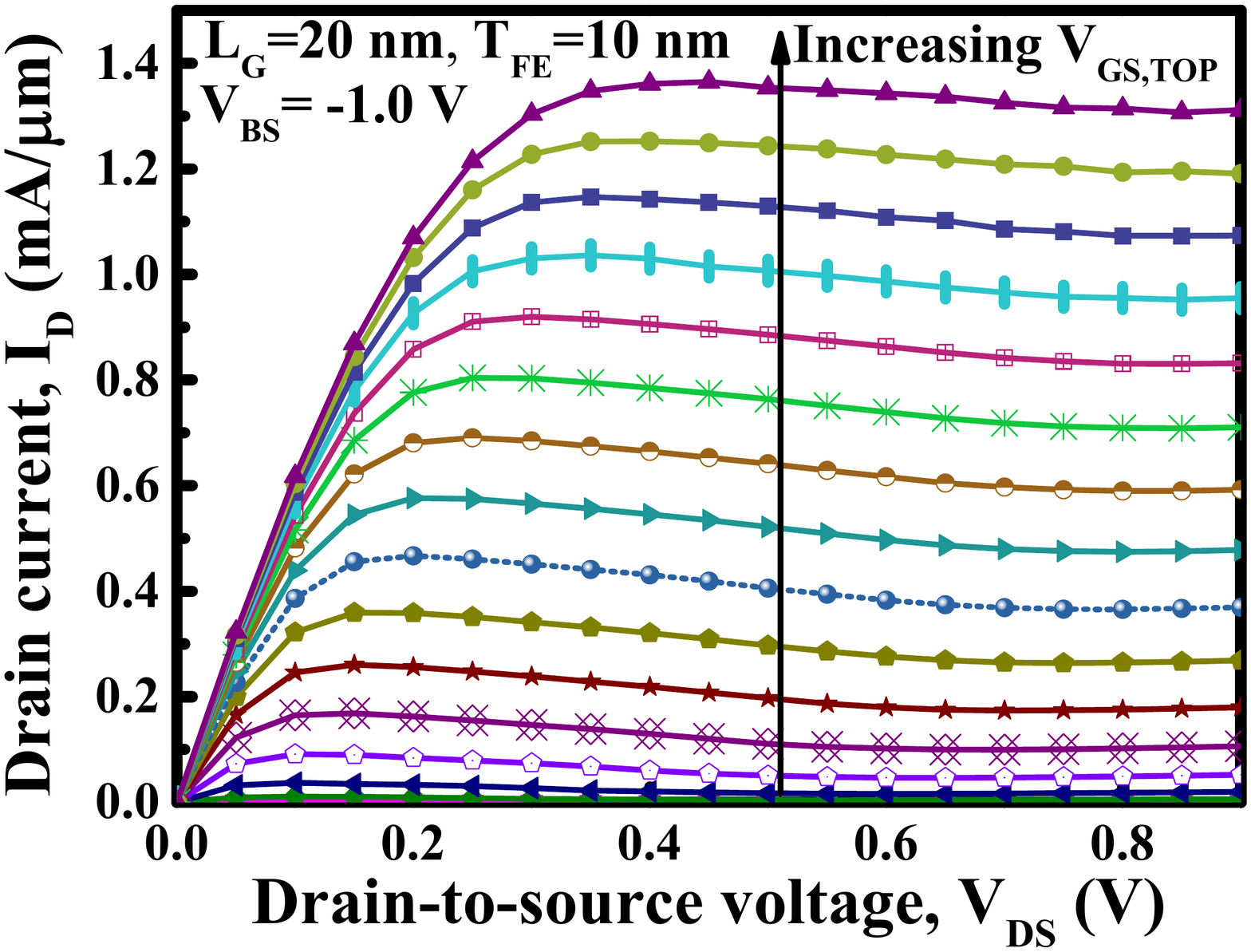}\label{n-ivd-hfo2}}
		 \subfigure[]{\includegraphics[trim=1cm 1.8cm 4cm 2cm clip=true,  width=4.25cm, height=4.0cm]{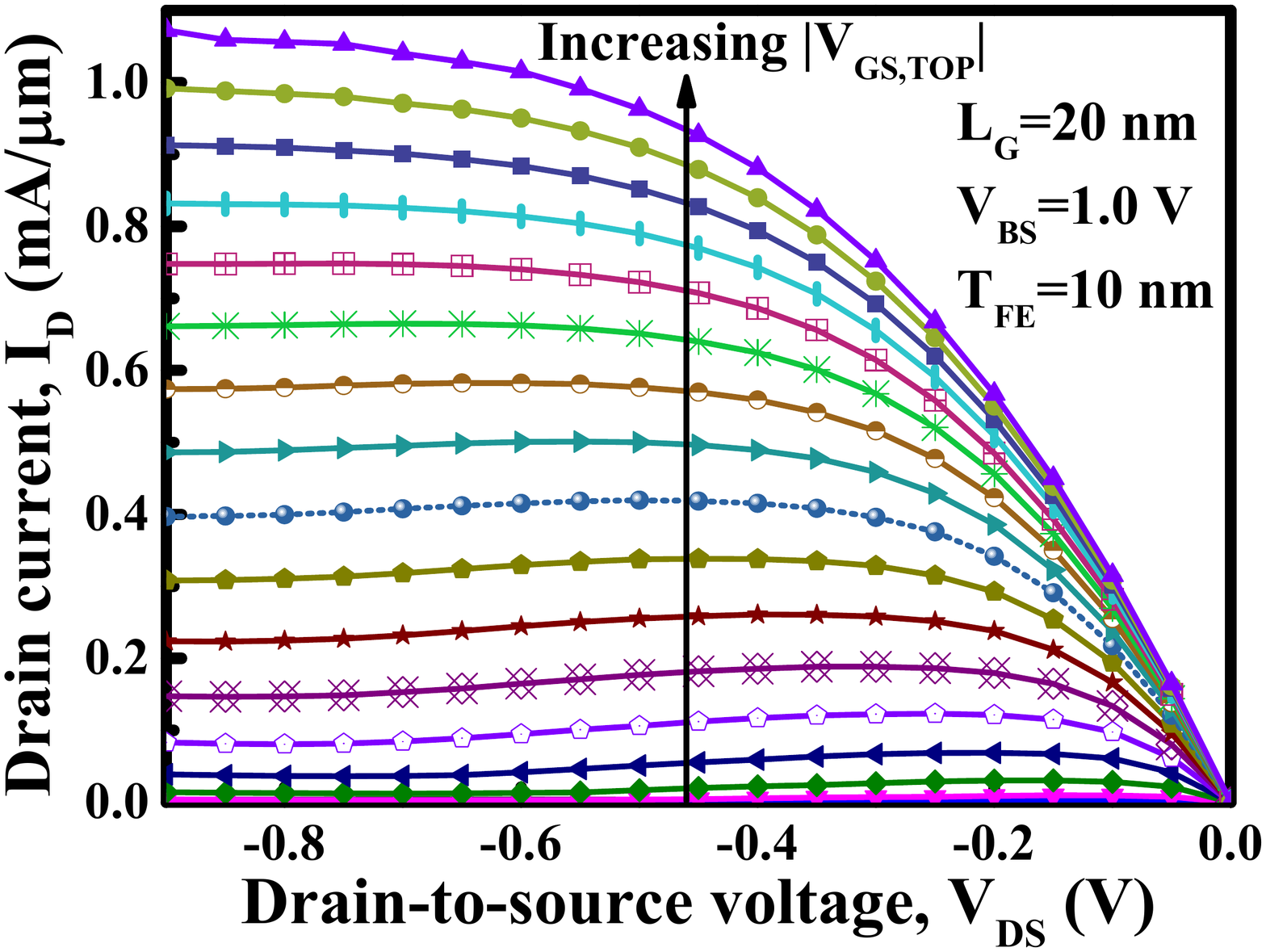}\label{p-ivd-hfo2}}
     \caption{I$_{D}$-V$_{DS}$ characteristics of FDSOI (a) n-NCFETs and, (b) p-NCFETs, with increasing V$_{GS,TOP}$, for T$_{FE}$=10 nm of HfO$_{2}$. $\mid$$V_{BS}$$\mid$=1 V.}
     \label{ivd_hfo2}
\end{figure}

\begin{figure}[!ht]
\setlength{\abovecaptionskip}{-3.5pt}
\vspace{-0.2in}
     \centering
     \subfigure[]{\includegraphics[trim=2cm 1.8cm 4cm 2cm clip=true, width=4.25cm,height=4.0cm]{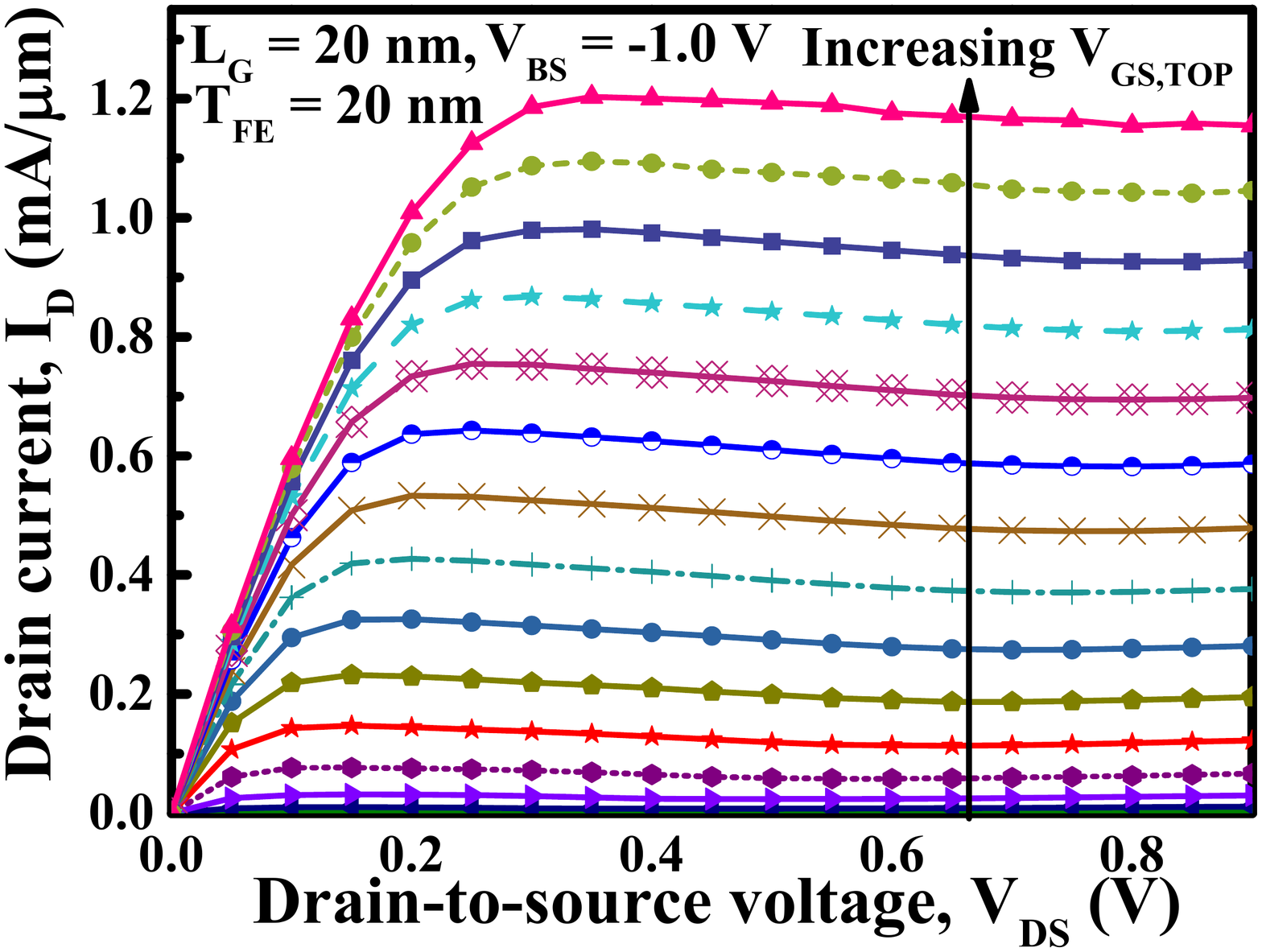}\label{3a}}
		 \subfigure[]{\includegraphics[trim=1cm 1.8cm 4cm 2cm clip=true, width=4.25cm,height=4.0cm]{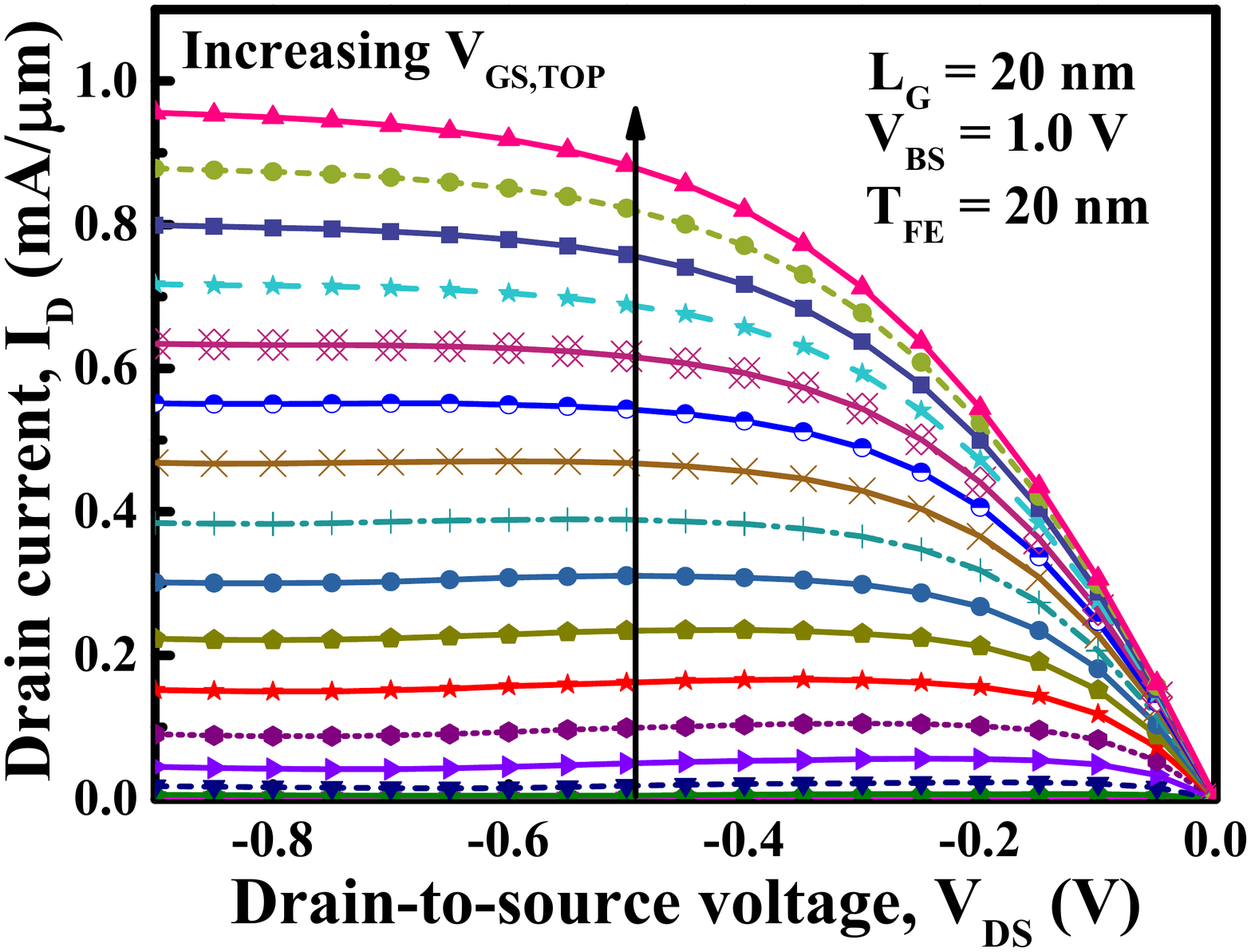}\label{3b}}
     \caption{I$_{D}$-V$_{DS}$ characteristics of FDSOI (a) n-NCFETs and, (b) p-NCFETs, with increasing V$_{GS,TOP}$, for T$_{FE}$=20 nm of PZT. $\mid$V$_{BS}$$\mid$=1 V.}
     \label{ivd_pzt}
\end{figure}

\subsection{Realization of PGP FDSOI NCFET}
The 2-D electrostatics obtained from TCAD were solved self consistently with Landau-Khalatnikov equation in MATLAB to simulate a PGP FDSOI NCFET \cite{Saeidi_TED_2016}. The Landau coefficients of HfO$_{2}$ ($\alpha$ = -3.9e10 cm/F, $\beta$ = 1.0e20 cm$^{5}$/F/C$^{2}$, $\gamma$ = -2.65e28 cm$^{9}$/F/C$^{4}$) were obtained from \cite{Esseni_EDL} and those of PZT ($\alpha$ = -13.5e9 cm/F, $\beta$ = 3.05e18 cm$^{5}$/F/C$^{2}$, $\gamma$ = -2.11e25 cm$^{9}$/F/C$^{4}$) were obtained from \cite{Saeidi_TED_2016}. FDSOI NCFETs used in our study with HfO$_{2}$ had a ferroelectric thickness (T$_{FE}$) of 10 nm. Even though a T$_{FE}$ of 20 nm for HfO$_{2}$ did not show hysteresis in DC simulations, as shown in Fig. \ref{ivg_hfo2}, hysteresis was observed at the circuit level for this T$_{FE}$. This is consistent with the behaviour reported in \cite{DIBR}. Similarly, as shown in Fig. \ref{ivg_pzt}, FDSOI NCFETs with a T$_{FE}$ of 30 nm for PZT did not show hysteresis in DC simulations. But hysteretic behaviour was observed for this value of T$_{FE}$ of PZT at the circuit level. Using constant current method, at a drain current of 100 $\mu$A/$\mu$m, the threshold voltage of HfO$_{2}$ FDSOI NCFETs (T$_{FE}$=10 nm) was 0.34 V (n-type) and -0.3 V (p-type). Similarly, the threshold voltage of PZT FDSOI NCFETs (T$_{FE}$ = 20 nm) was 0.43 V (n-type) and -0.42 V (p-type). The output characteristics of HfO$_{2}$ FDSOI NCFETs (T$_{FE}$ = 10 nm) are shown in Fig. \ref{ivd_hfo2} and the output characteristics of PZT FDSOI NCFETs (T$_{FE}$ = 20 nm) are shown in Fig. \ref{ivd_pzt}. The negative DIBL behaviour at high V$_{GS,TOP}$ in FDSOI p-NCFET is smaller than in FDSOI n-NCFET and is consistent with the observation and explanation provided in \cite{NDR_FinFET_TED2017}. The increased thickness of PZT in comparison to HfO$_{2}$ is consistent with our reported study based on $\alpha$, $\beta$, $\gamma$ coefficients which showed that HfO$_{2}$ ferroelectric is expected to give greater non-hysteretic gain in comparison to PZT for given T$_{FE}$\cite{2018EDTM}.

\section{Evaluation of Logic Gates using NCFETs}
\label{sec:guidelines}
Two of the most popular ferroelectrics, HfO$_{2}$ and PZT have been used to study FDSOI NCFETs for digital circuits. The circuits studied were 3-stage CMOS ring oscillator, NAND-2 with fan-out of 1 and NOR-2 with fan-out of 1 \cite{2018EUROSOI}.  The circuits were simulated in Synopsys HSPICE using a look-up table approach \cite{HSPICE}. The circuits with FDSOI NCFETs were studied at a supply voltage (V$_{DD}$) of 0.5 V which is also the threshold voltage of baseline FDSOI MOSFETs. Further, the comparison of FDSOI NCFETs with baseline FDSOI MOSFETs was also drawn at V$_{DD}$ ranging from 0.5 V to 0.9 V. The propagation delay, $\tau_p$ was based on the charge-current relationship \cite{Rabaey}.

\begin{figure}[!ht]
\vspace{-0.3in}
\setlength{\abovecaptionskip}{-2.5pt}
\centerline{\includegraphics[width=\columnwidth]{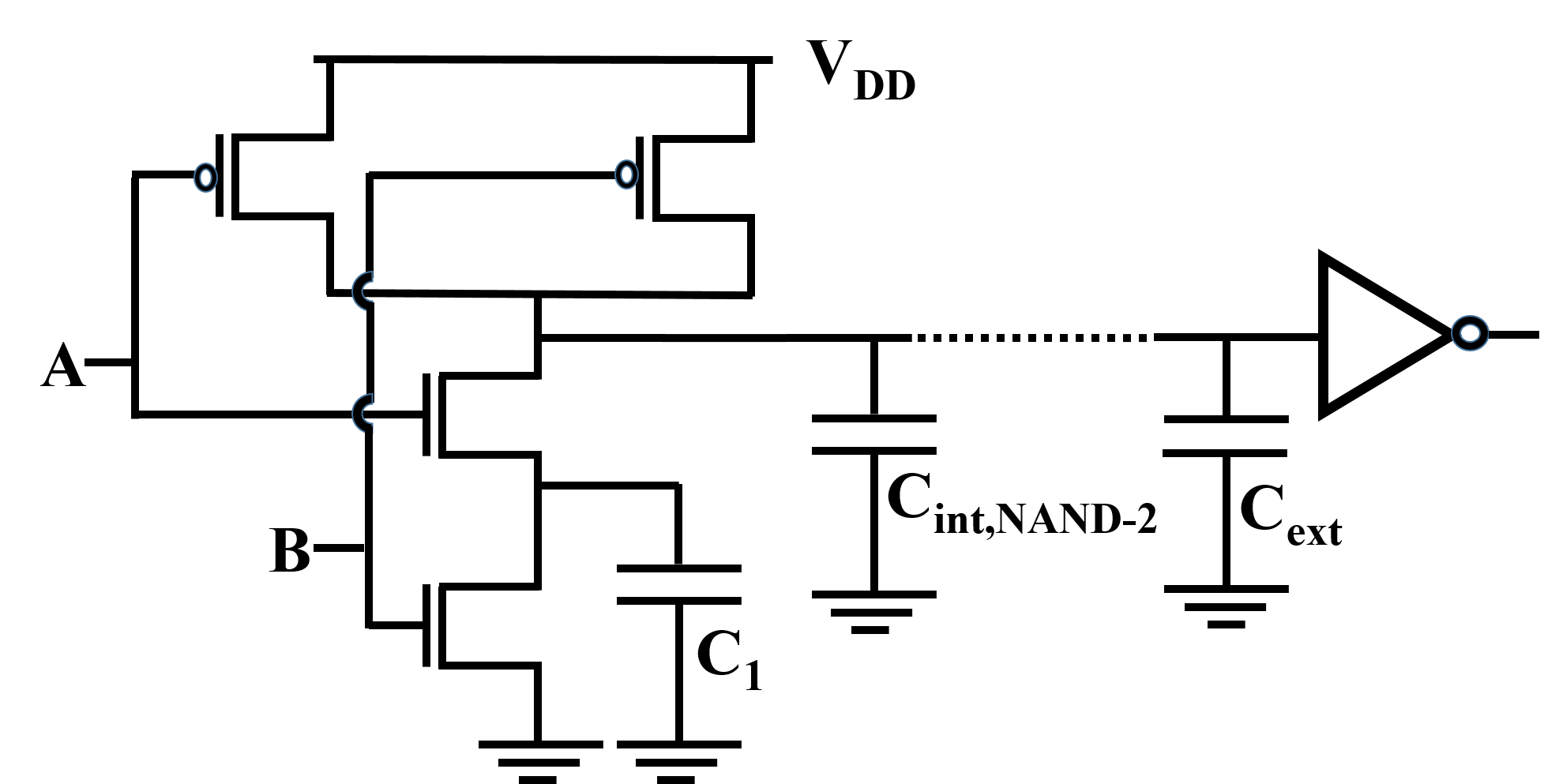}}
\caption{Schematic of NAND-2 gate followed by an inverter showing the intrinsic and external capacitances of NAND-2 which have been considered to evaluate the performance of the gate.}
\label{nand-2_schematic}
\end{figure}

\begin{figure}[!ht]
\setlength{\abovecaptionskip}{-2.5pt}
\centerline{\includegraphics[width=\columnwidth]{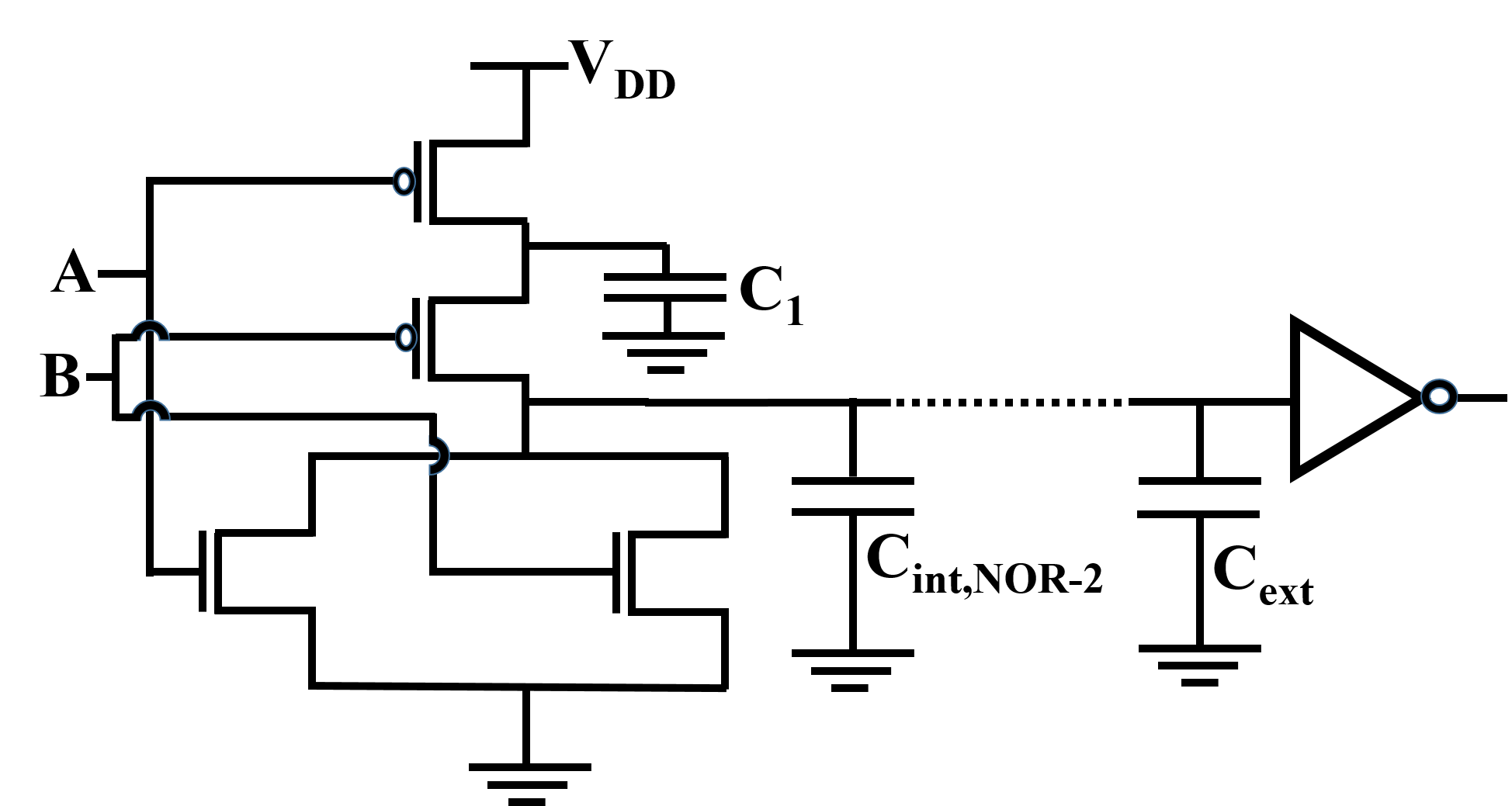}}
\caption{Schematic of NOR-2 gate followed by an inverter showing the intrinsic and external capacitances of NOR-2 which have been considered to evaluate the performance of the gate.}
\label{nor2_schematic}
\end{figure}

In this study, the p-type devices were not scaled and in case of both baseline FDSOI MOSFETs and FDSOI NCFETs, the n-type and p-type devices had the same dimensions. The load capacitance (C$_{load}$) at the output of each inverter in the ring oscillator was taken as the sum of output capacitance of the previous stage and the input capacitance of the next stage. To obtain gate delay, a model assumption was made that the intrinsic capacitance, C$_{int}$, of the logic gate scales with the gate capacitance of the device (C$_{G}$), i.e., C$_{int}$ $\simeq$ C$_{G}$. The external capacitance (C$_{ext}$) in case of the baseline device was taken as the input capacitance of the inverter that loads the gate, which is, C$_{ext}$ = C$_{GS,n}$ +  C$_{GS,p}$ + C$_{GD,n}$(1-A$_{V}$) + C$_{GD,p}$(1-A$_{V}$), as shown in Figs. \ref{nand-2_schematic} and \ref{nor2_schematic}. A similar expression was dervied for the case of FDSOI NCFETs taking the ferroelectric capacitance, C$_{FE}$ into account. The capacitances C$_{GS,n}$, C$_{GS,p}$, C$_{GD,n}$ and C$_{GD,p}$ for the baseline device were obtained from TCAD and C$_{FE}$ was calculated as C$_{FE}$ $\approx$ 1/(2$\alpha$T$_{FE}$) as outlined in \cite{CFE}. Here, A$_{V}$ is the gain of the inverter (=-1). For case of both baseline FDSOI MOSFETs and FDSOI NCFETs based gates, C$_{ext}$ is approximately equal to C$_{G}$.  For logic gate, C$_{load}$ = C$_{G}$ + C$_{ext}$. The input capacitance of the gate (C$_{in}$) was 2C$_{G}$. Therefore, for a fan-out of 1, C$_{load}$/C$_{in}$ $\simeq$ 1, is verified. The capacitance C$_{1}$ = C$_{DB}$ + C$_{SB}$ + C$_{GD}$(1-A$_{V}$).  Miller effect was considered in our analysis. Worst-case input patterns were considered for obtaining gate delays. For NAND-2 gate, the worst case input pattern was taken as A=1, B=0$\rightarrow$1 for $\tau_{p,HL}$ and A=1, B=1$\rightarrow$0 for $\tau_{p,LH}$. Similarly, for NOR-2, the worst case input pattern for $\tau_{p,HL}$ was taken as A=0$\rightarrow$1, B=0 and A=1$\rightarrow$0, B=0 for $\tau_{p,LH}$.

\begin{table}
\centering
\setlength{\tabcolsep}{3pt}
\caption{Performance of 3-stage ring oscillators, $\mid$V$_{BS}$$\mid$=1 V.}
\label{ro_table}
\begin{tabular}{|c|c|c|c|}
\hline
\textbf{V$_{DD}$ (V)} & \textbf{\textit{f$_{osc,BL}$} (GHz)} & \textbf{\textit{f$_{osc,PZT}$} (GHz)} & \textbf{\textit{f$_{osc,HfO_{2}}$} (GHz)}\\
\hline
0.5 & 21 & 53 & 67\\
\hline
0.6 & 46 & 84 & 101\\
\hline
0.7 & 73 & 113 & 126\\
\hline
0.8 & 99 & 140 & 151\\
\hline
0.9 & 120 & 162 & 171\\
\hline
\end{tabular}

\end{table}

\begin{figure}[!ht]
\setlength{\abovecaptionskip}{-3pt}
\centerline{\includegraphics[trim=2cm 1cm 2.5cm 1.8cm clip=true, width=7.5cm]{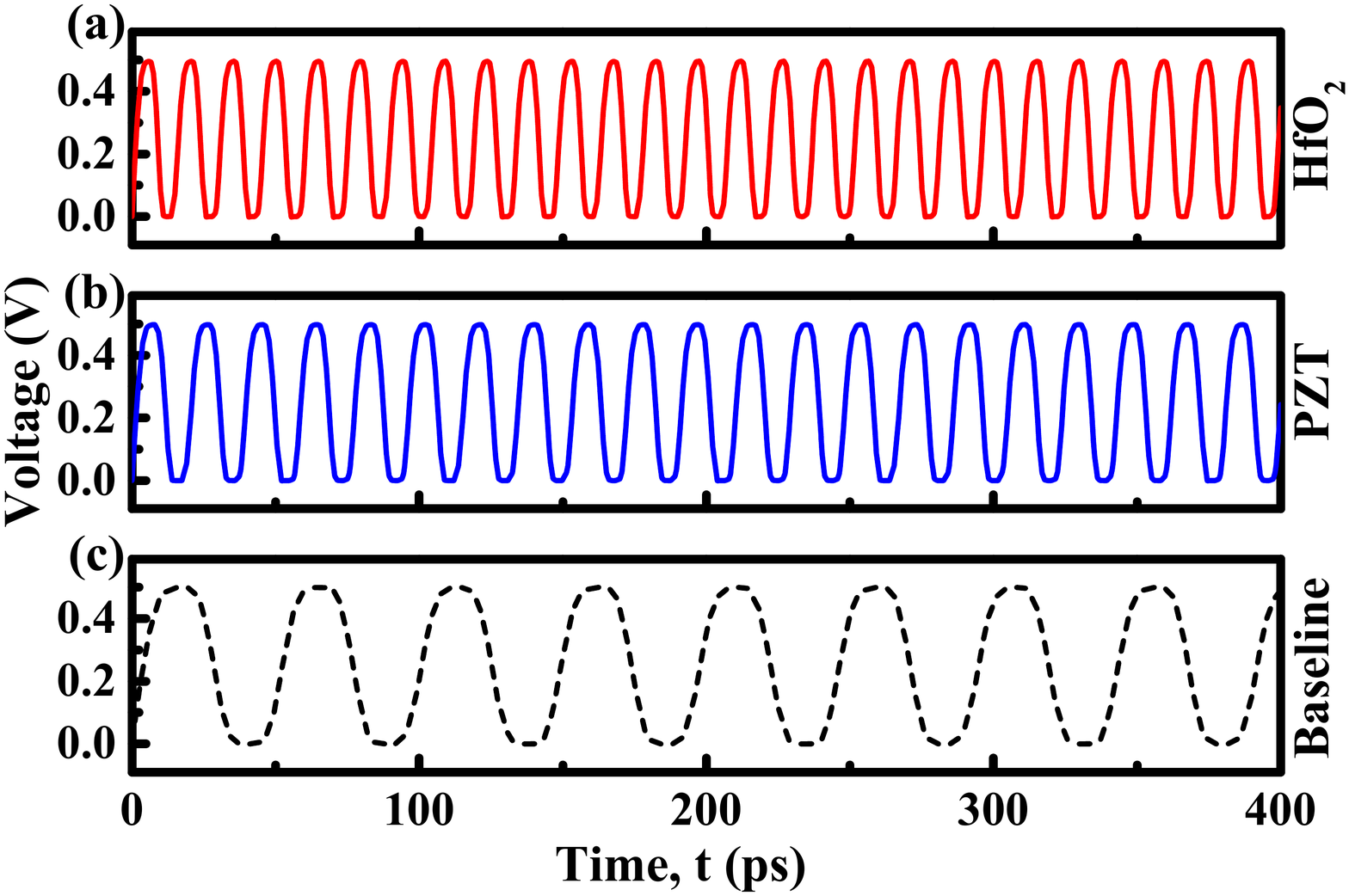}}
\caption{Performance of 3-stage ring oscillator using (a) HfO$_{2}$ FDSOI NCFETs (\textit{f$_{osc}$}=67 GHz), (b) PZT FDSOI NCFETs (\textit{f$_{osc}$}=53 GHz) and, (c) baseline FDSOI MOSFETs (\textit{f$_{osc}$}=21 GHz). V$_{DD}$ is 0.5 V, $\mid$V$_{BS}$$\mid$=1 V. }
\label{ro_graph}
\end{figure}

The results obtained for frequency of oscillation (\textit{f$_{osc}$}) of ring oscillators are shown in Table \ref{ro_table}. For baseline FDSOI MOSFET based ring oscillator, \textit{ f$_{osc}$} at V$_{DD}$ of 0.5 V was 21 GHz while HfO$_{2}$ FDSOI NCFET based ring oscillator had \textit{ f$_{osc}$} of 67 GHz at the same V$_{DD}$, as shown in Fig. \ref{ro_graph}. PZT FDSOI NCFET based ring oscillator had \textit{f$_{osc}$} of 53 GHz which was greater than in case of the baseline device based ring oscillator but less than that of HfO$_{2}$ FDSOI NCFET based ring oscillator. For analyzing the performance and average power consumption of the logic gates, the input signal frequency was varied from 5 KHz to 20 GHz, which is consistent with results in Table \ref{ro_table}.

\begin{figure}[!ht]
\setlength{\abovecaptionskip}{-3pt}
\centerline{\includegraphics[trim=2cm 1cm 0.5cm 2cm clip=true, width=0.8\columnwidth]{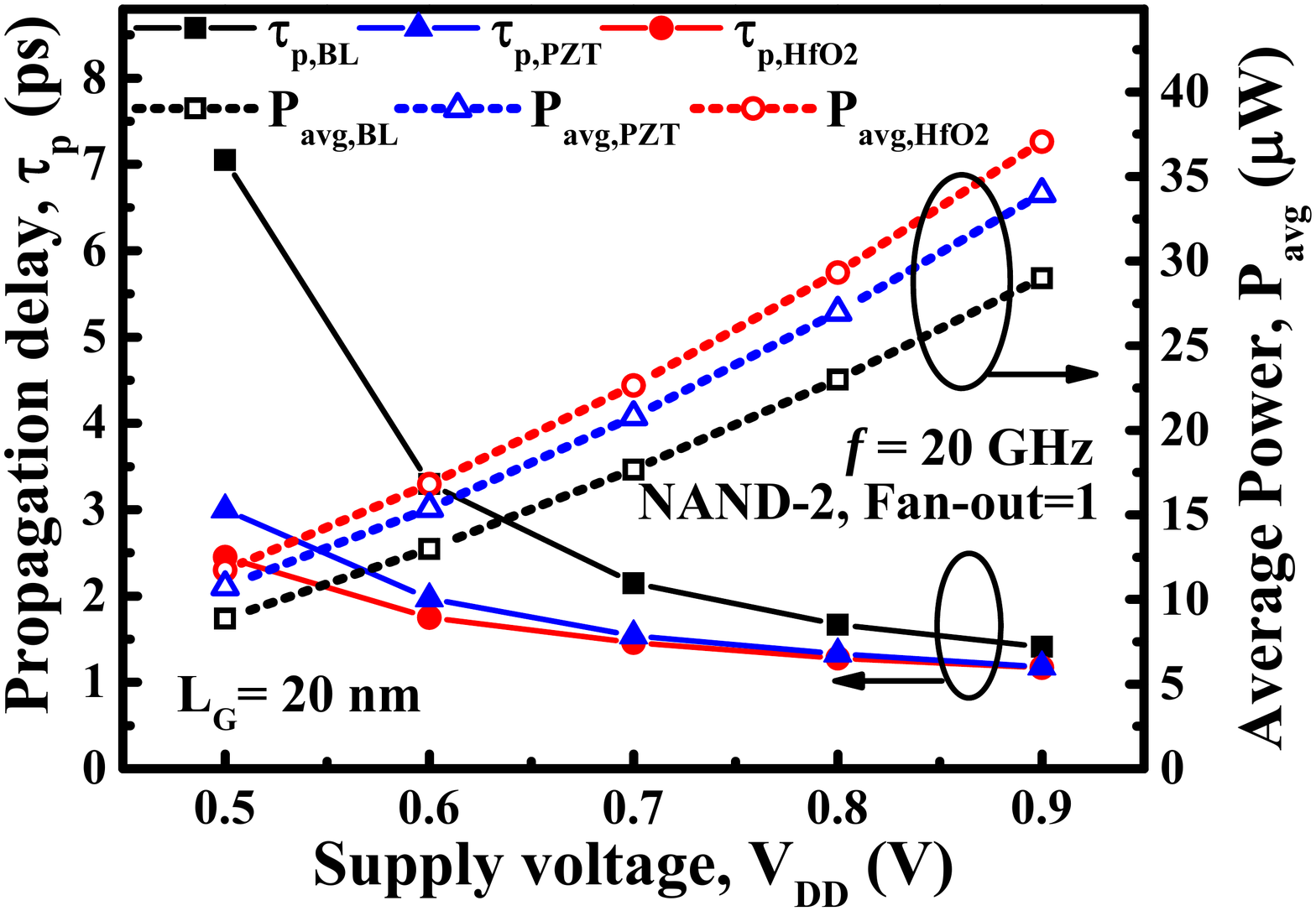}}
\caption{Propagation delay and average power consumption comparison at different V$_{DD}$ for HfO$_{2}$ FDSOI NCFET based NAND-2 gate, PZT FDSOI NCFET based NAND-2 gate and FDSOI MOSFET based NAND-2 gate. Fan-out is 1, $\mid$V$_{BS}$$\mid$=1 V.}
\label{nand-2_graph}
\vspace{-0.1in}
\end{figure}

The results obtained for performance and average power consumption for NAND-2 gates with fan-out of 1 are shown in Tabel \ref{nand2_table} at different V$_{DD}$. It can be clearly seen that for comparable performance at 20 GHz of baseline FDSOI MOSFET based NAND-2 gate (V$_{DD}$ = 0.7 V), PZT FDSOI NCFET based NAND-2 gate (V$_{DD}$ = 0.6 V) and HfO$_{2}$ FDSOI NCFET based NAND-2 gate (V$_{DD}$ = 0.5 V), the average power consumption is least in HfO$_{2}$ FDSOI NCFET based NAND-2 gate. For HfO$_{2}$ FDSOI NCFET based NAND-2 gate, the average power consumption was $\sim$66\% and for PZT FDSOI NCFET based NAND-2 gate it was $\sim$86\% that of baseline FDSOI MOSFET based NAND-2 gate. This holds true for other V$_{DD}$ values also as shown in Table \ref{nand2_table}. This finding is consistent with the analysis reported in \cite{Harshit_EDL_2018}. Figure \ref{nand-2_graph} shows the variation of propagation delay and average power consumption of NAND-2 gate for all cases at different V$_{DD}$. 

\begin{table}[!ht]
\centering
\setlength{\tabcolsep}{3pt}
\caption{Propagation delay and average power consumption for NAND-2 gates using FDSOI NCFETs and baseline devices. The signal frequency is 20 GHz, $\mid$V$_{BS}$$\mid$=1 V.}
\label{nand2_table}
\begin{tabular}{|c|c|c|c|c|c|c|}
\hline
\textbf{V$_{DD}$} & \textbf{$\tau_{p,BL}$} & \textbf{$\tau_{p,PZT}$} & \textbf{$\tau_{p,HfO_{2}}$} & \textbf{P$_{avg,BL}$} & \textbf{P$_{avg,PZT}$} & \textbf{P$_{avg,HfO_{2}}$}\\
\textbf{(V)}      & \textbf{(ps)}   & \textbf{ps}              & \textbf{(ps)}  & \textbf{($\mu$W)}     & \textbf{($\mu$W)} & \textbf{($\mu$W)}\\
\hline
0.5 & 7.05 & 3 & \textcolor{red}{2.45} & 8.86 & 10.75 & \textcolor{red}{11.74}\\
\hline
0.6 & 3.3 & \textcolor{red}{1.97} & \textcolor{blue}{1.75} & 13 & \textcolor{red}{15.38} & \textcolor{blue}{16.82}\\
\hline
0.7 & \textcolor{red}{2.15} & \textcolor{blue}{1.54} & \textcolor{magenta}{1.46} & \textcolor{red}{17.69} & \textcolor{blue}{20.8} & \textcolor{magenta}{22.65}\\
\hline
0.8 & \textcolor{blue}{1.67} & \textcolor{magenta}{1.33} & 1.28 & \textcolor{blue}{23} & \textcolor{magenta}{26.98} & 29.32\\
\hline
0.9 & \textcolor{magenta}{1.41} & 1.18 & 1.17 & \textcolor{magenta}{29} & 33.99 & 37.06\\
\hline
\end{tabular}
\end{table}

\begin{table}[!ht]
\centering
\setlength{\tabcolsep}{3pt}
\caption{Propagation delay and average power consumption for NOR-2 gates using FDSOI NCFETs and baseline devices. The signal frequency is 20 GHz, $\mid$V$_{BS}$$\mid$=1 V.}
\label{nor2_table}
\begin{tabular}{|c|c|c|c|c|c|c|}
\hline
\textbf{V$_{DD}$} & \textbf{$\tau_{p,BL}$} & \textbf{$\tau_{p,PZT}$} & \textbf{$\tau_{p,HfO_{2}}$} & \textbf{P$_{avg,BL}$} & \textbf{P$_{avg,PZT}$} & \textbf{P$_{avg,HfO_{2}}$}\\
\textbf{(V)}      & \textbf{(ps)}   & \textbf{ps}              & \textbf{(ps)}  & \textbf{($\mu$W)}     & \textbf{($\mu$W)} & \textbf{($\mu$W)}\\
\hline
0.5 & - & 3.55 & 2.96 & - & 10.68 & 11.63\\
\hline
0.6 & 3.8 & 2.3 &  \textcolor{red}{2.08} & 13 & 15.18 &  \textcolor{red}{16.52}\\
\hline
0.7 & 2.5 &  \textcolor{red}{1.77} & \textcolor{blue}{1.68} & 17.7 &  \textcolor{red}{20.46} & \textcolor{blue}{22.27}\\
\hline
0.8 &  \textcolor{red}{1.9} & \textcolor{blue}{1.5} & 1.46 &  \textcolor{red}{23} & \textcolor{blue}{26.47} & 28.8\\
\hline
0.9 &  \textcolor{blue}{1.6} & 1.33 & 1.32 & \textcolor{blue}{29} & 33.26 & 36.24\\
\hline
\end{tabular}
\end{table}

\begin{figure}[!ht]
\setlength{\abovecaptionskip}{-3pt}
\centerline{\includegraphics[trim=2cm 1cm 0.5cm 2cm clip=true, width=0.8\columnwidth]{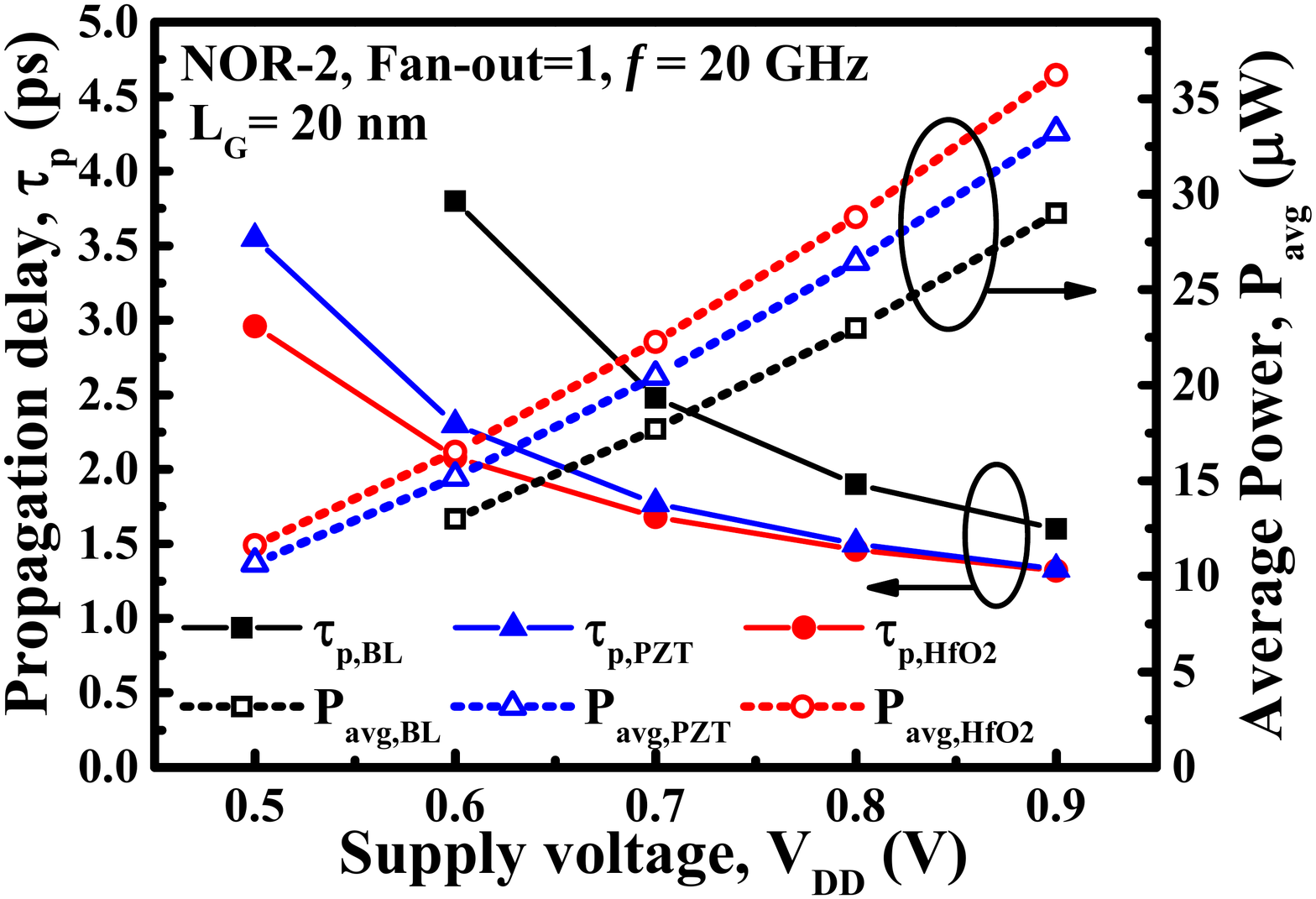}}
\caption{Propagation delay and average power consumption comparison at different V$_{DD}$ for HfO$_{2}$ FDSOI NCFET based NOR-2 gate, PZT FDSOI NCFET based NOR-2 gate and FDSOI MOSFET based NOR-2 gate. The baseline FDSOI MOSFET based NOR-2 gate failed to function at 20 GHz at a V$_{DD}$ of 0.5 V. Fan-out is 1, $\mid$V$_{BS}$$\mid$=1 V.}
\label{nor2_graph}
\vspace{-0.2in}
\end{figure}

Table \ref{nor2_table} shows the results obtained for performance and average power consumption for NOR-2 gates. The baseline FDSOI MOSFET based NOR-2 gates could not be operated beyond 10 GHz at a V$_{DD}$ of 0.5 V. At comparable performance of $\sim$2 ps at 20 GHz, the average power consumption of HfO$_{2}$ FDSOI NCFET based NOR-2 gate was $\sim$72\% and for PZT FDSOI NCFET based NOR-2 gate it was $\sim$88\% that of baseline FDSOI MOSFET based NOR-2 gate. The improved performance with reduced power consumption for HfO$_{2}$ FDSOI NCFETs based NOR-2 gate is shown in Fig. \ref{nor2_graph}.

\begin{figure}[t]
\setlength{\abovecaptionskip}{-3pt}
\centerline{\includegraphics[trim=2cm 1cm 0.5cm 2cm clip=true, width=0.8\columnwidth]{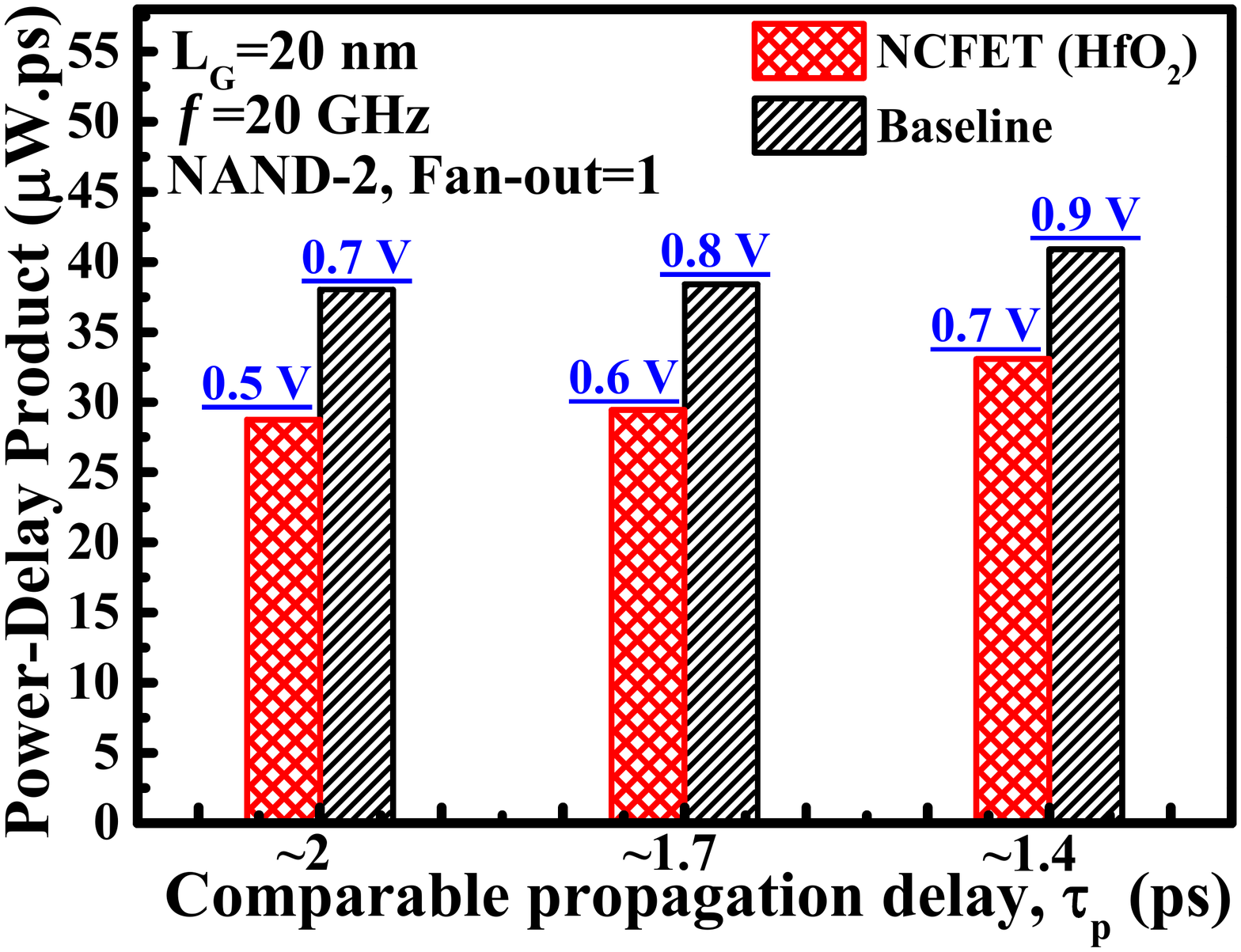}}
\caption{Power-delay product of HfO$_{2}$ FDSOI NCFET based NAND-2 gates and FDSOI MOSFET based NAND-2 gates with fan-out of 1 at 20 GHz, when the gates have comparable performance for minimum sized transistors. The underlined numbers represent V$_{DD}$ values. $\mid$V$_{BS}$$\mid$=1 V.}
\label{pdp_nand2_hfo2}
\end{figure}

\begin{figure}[h]
\setlength{\abovecaptionskip}{-3pt}
\centerline{\includegraphics[trim=2cm 1cm 0.5cm 2cm clip=true,, width=0.8\columnwidth]{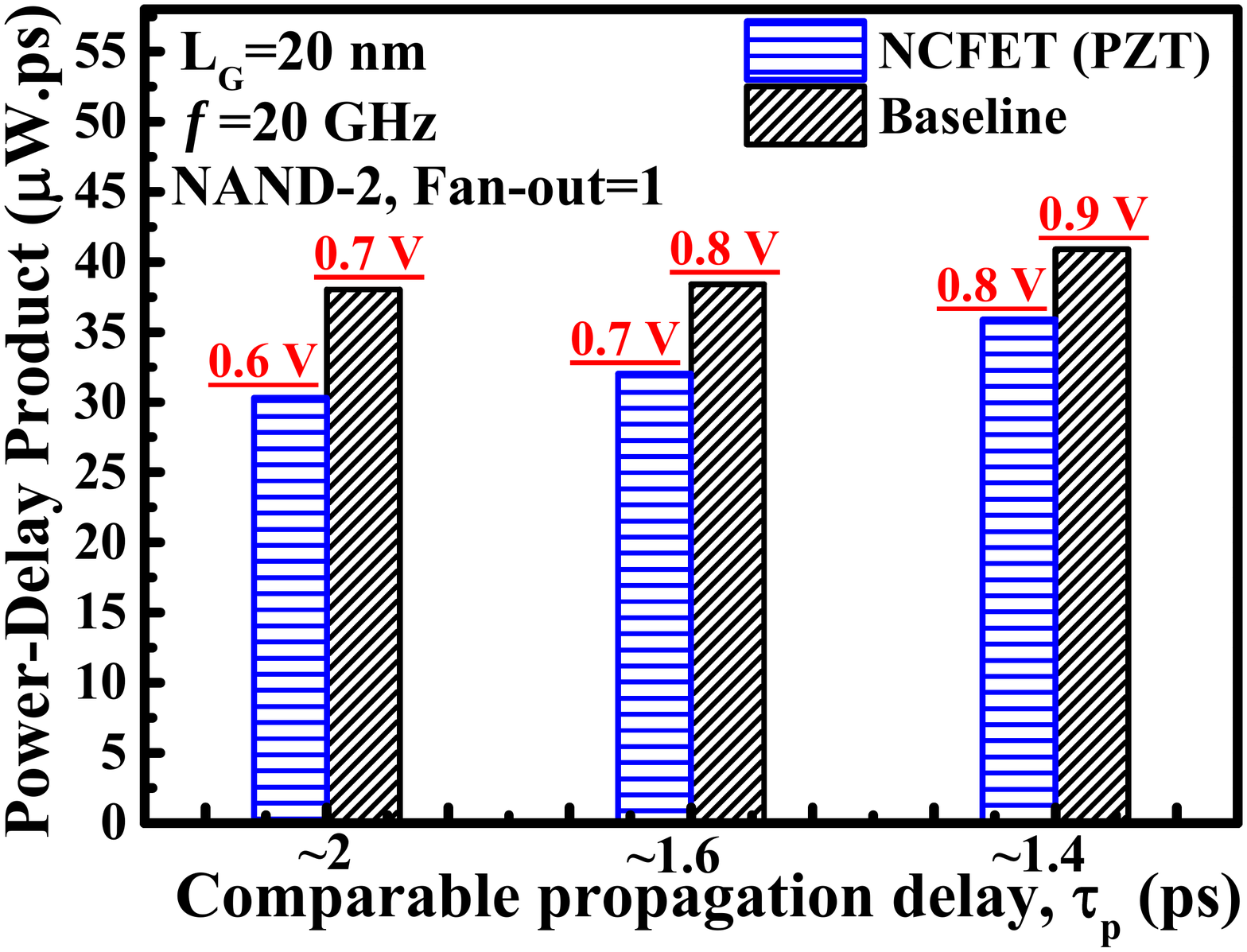}}
\caption{Power-delay product of PZT FDSOI NCFET and FDSOI MOSFET based NAND-2 gates with fan-out of 1 at 20 GHz, when the gates have comparable performance for minimum sized transistors. The underlined numbers represent V$_{DD}$ values. $\mid$V$_{BS}$$\mid$=1 V.}
\label{pdp_nand2_pzt}
\end{figure}

\subsection{Power-Delay Product}
Power-Delay Product (PDP) is an important metric for digital circuits and gives an estimate of the amount of energy consumed in an operation. The reduction in PDP of FDSOI NCFETs based NAND-2 gates over those with FDSOI MOSFET based NAND-2 gates is shown in Figs. \ref{pdp_nand2_hfo2} and \ref{pdp_nand2_pzt} for HfO$_{2}$ and PZT ferroelectrics, respectively, at signal frequency of 20 GHz. For a comparable propagation delay of $\sim$2 ps, HfO$_{2}$ FDSOI NCFET based NAND-2 gates had a PDP of $\sim$29 $\mu$W.ps (V$_{DD}$ = 0.5 V), PZT FDSOI NCFET based NAND-2 gates had a PDP of $\sim$30.3 $\mu$W.ps (V$_{DD}$ = 0.6 V) while baseline FDSOI MOSFET based NAND-2 gates had a PDP of $\sim$38 $\mu$W.ps (V$_{DD}$ = 0.7 V). 

\begin{figure}[!ht]
\setlength{\abovecaptionskip}{-3pt}
\centerline{\includegraphics[trim=2cm 1cm 0.5cm 2cm clip=true, width=0.8\columnwidth]{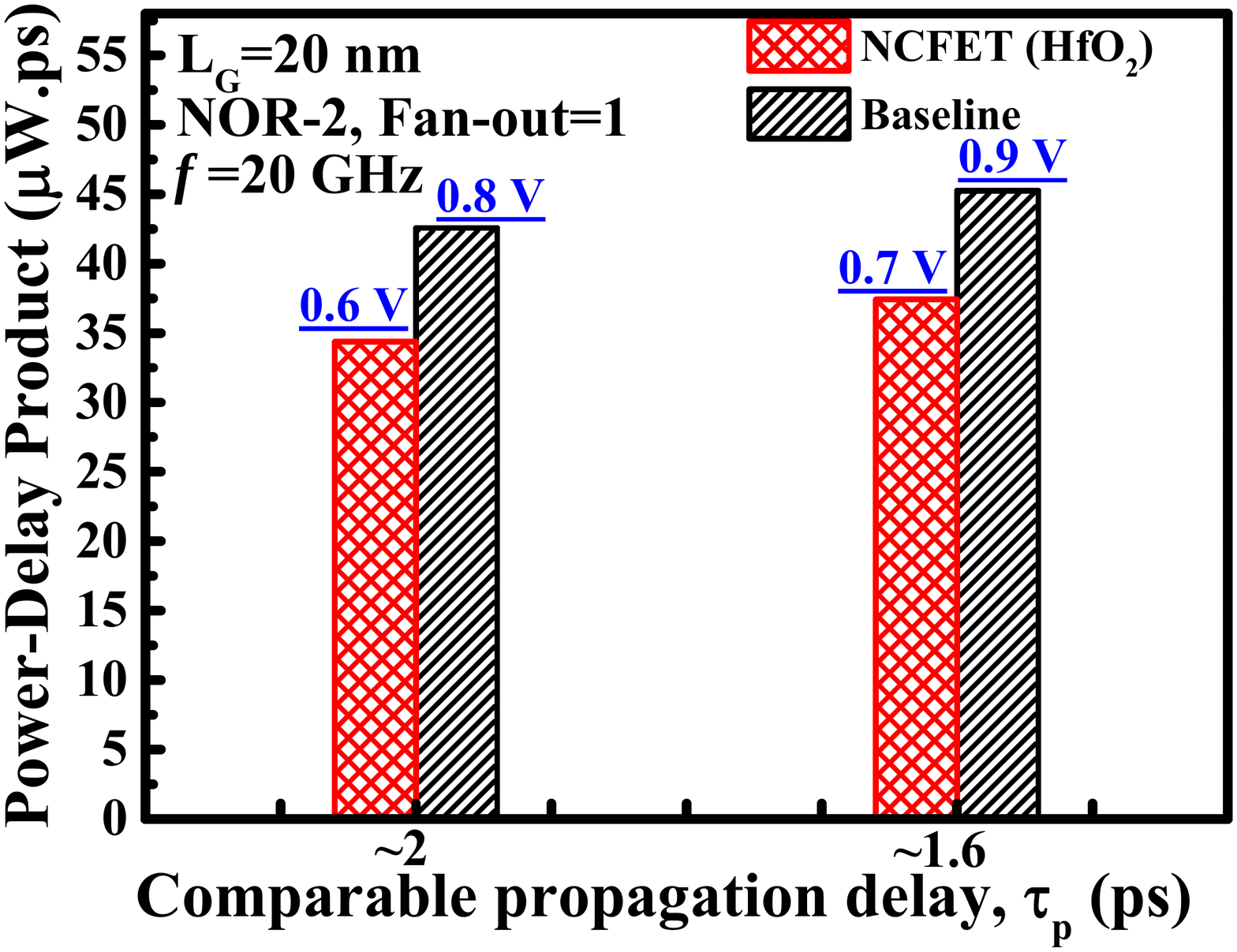}}
\caption{Power-delay product of HfO$_{2}$ FDSOI NCFET and FDSOI MOSFET based NOR-2 gates with fan-out of 1 at 20 GHz, when the gates have comparable performance for minimum sized transistors. The underlined numbers represent V$_{DD}$ values. $\mid$V$_{BS}$$\mid$=1 V.}
\label{pdp_nor2_hfo2}
\end{figure}

\begin{figure}[!ht]
\setlength{\abovecaptionskip}{-3pt}
\centerline{\includegraphics[trim=2cm 1cm 0.5cm 2cm clip=true, width=0.8\columnwidth]{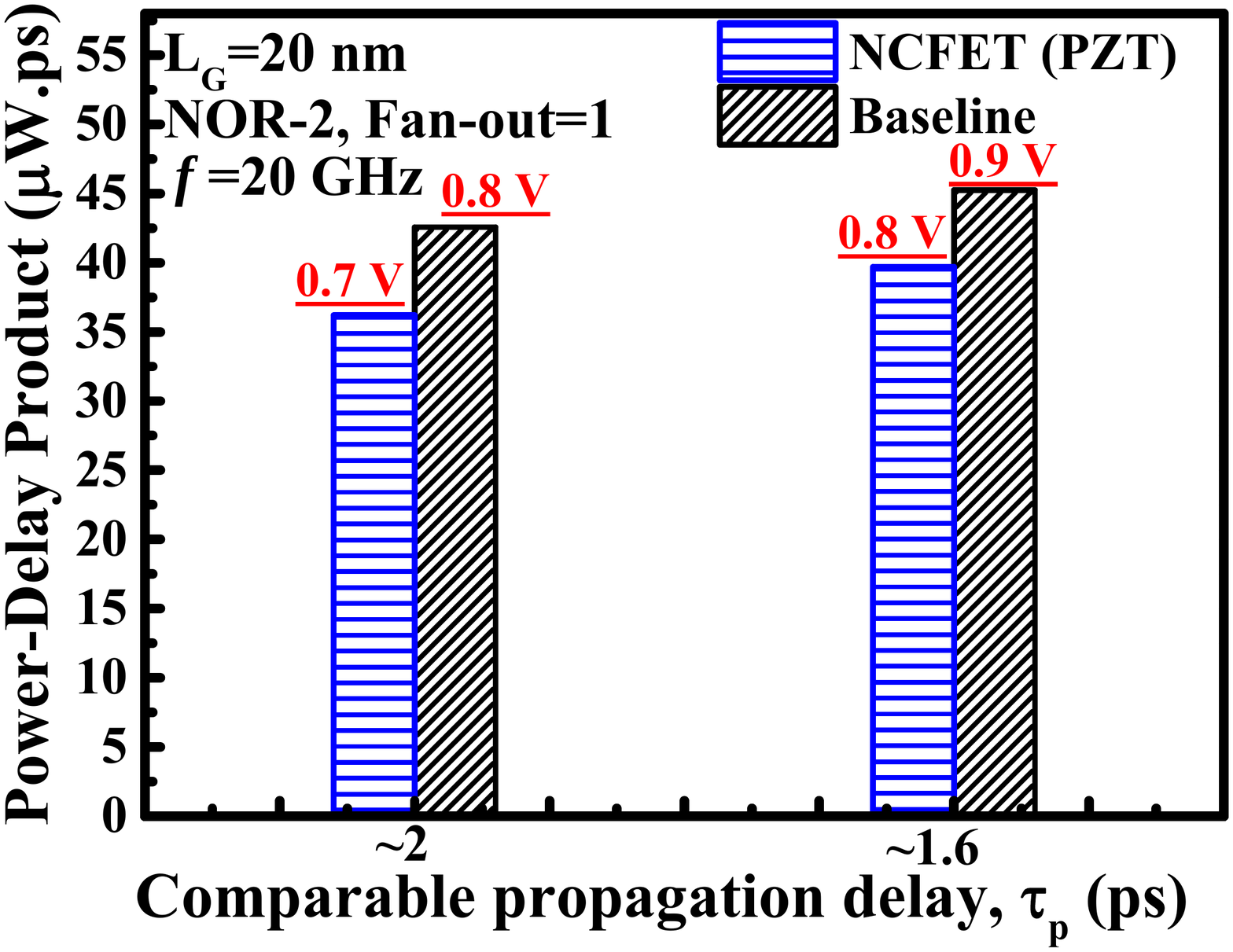}}
\caption{Power-delay product of PZT FDSOI NCFET and FDSOI MOSFET based NOR-2 gates with fan-out of 1 at 20 GHz, when the gates have comparable performance for minimum sized transistors. The underlined numbers represent V$_{DD}$ values. $\mid$V$_{BS}$$\mid$=1 V.}
\label{pdp_nor2_pzt}
\vspace{-0.1in}
\end{figure}

The PDP results for HfO$_{2}$ FDSOI NCFET based NOR-2 gate are shown in Fig. \ref{pdp_nor2_hfo2} and for PZT FDSOI NCFET based NOR-2 gate the results are shown in Fig. \ref{pdp_nor2_pzt}. The PDP in case of baseline FDSOI MOSFET based NOR-2 gates for a propagation delay of $\sim$2 ps at 20 GHz, was nearly $\sim$43 $\mu$W.ps at V$_{DD}$ of 0.8 V. For the same delay, the PDP of PZT FDSOI NCFET based NOR-2 gate was $\sim$36 $\mu$W.ps at a V$_{DD}$ of 0.7 V and PDP for HfO$_{2}$ FDSOI NCFET based NOR-2 gate was $\sim$34.3 $\mu$W.ps at a V$_{DD}$ of 0.6 V. The PDP results clearly highlight the significance of the thin ($\sim$10 nm) layer of HfO$_{2}$ ferroelectric in FDSOI NCFETs for digital circuit design. The analysis shows that thin layers of ferroelectrics in FDSOI NCFETs help in the improvement of device performance for digital circuit applications.

\subsection{Effect of fan-in and fan-out}
\begin{figure}[!ht]
\setlength{\abovecaptionskip}{-3pt}
\centerline{\includegraphics[trim=2cm 1cm 0.5cm 2cm clip=true,width=0.8\columnwidth]{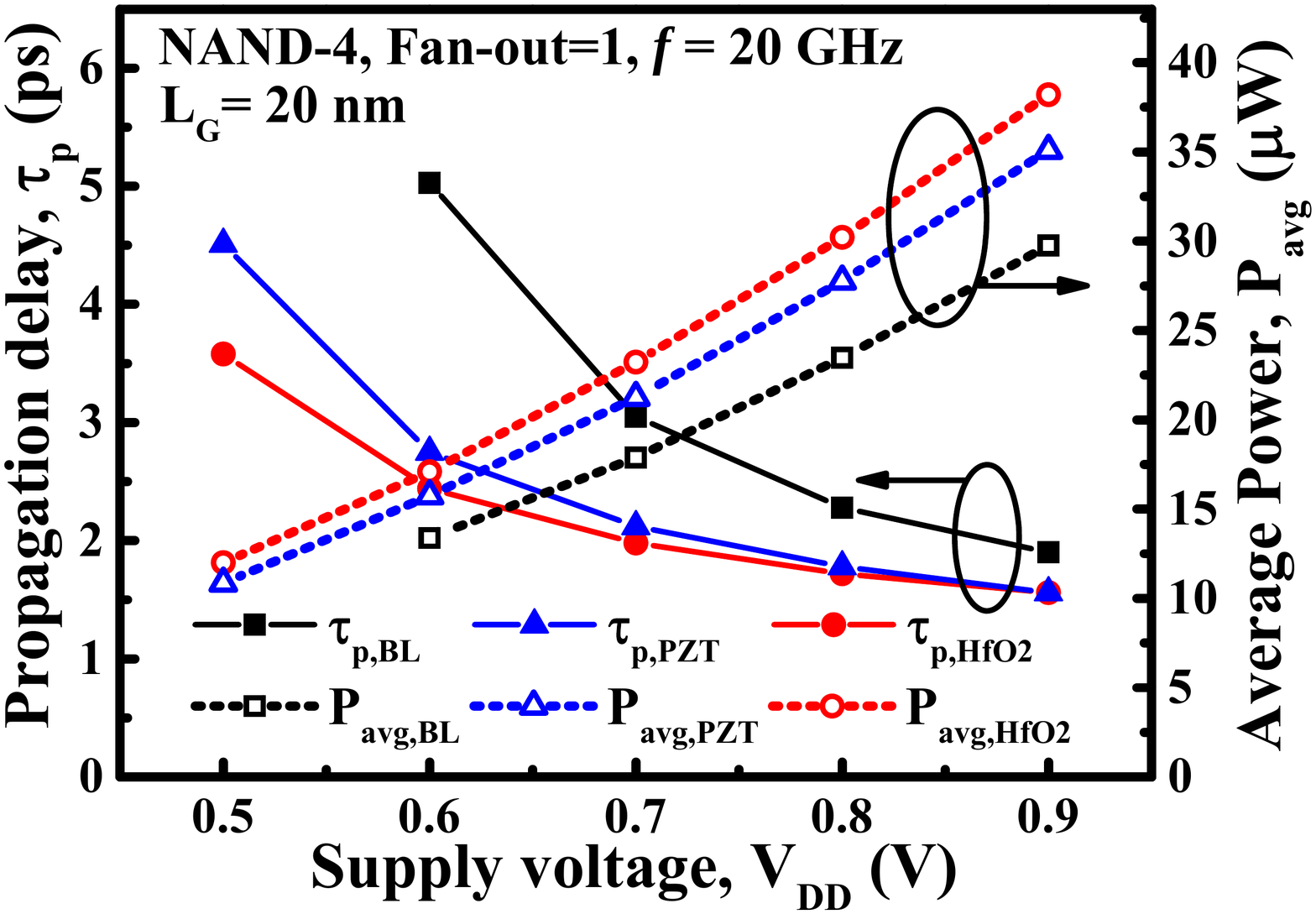}}
\caption{Propagation delay and average power consumption comparison at different V$_{DD}$ for HfO$_{2}$ FDSOI NCFET based NAND-4 gate, PZT FDSOI NCFET based NAND-4 gate and FDSOI MOSFET based NAND-4 gate. The baseline FDSOI MOSFET based NAND-4 gate failed to function at 20 GHz at a V$_{DD}$ of 0.5 V. Fan-out is 1, $\mid$V$_{BS}$$\mid$=1 V.}
\label{nand4_graph}
\end{figure}

\begin{figure}[!h]
\setlength{\abovecaptionskip}{-3pt}
\centerline{\includegraphics[trim=2cm 1cm 0.5cm 2cm clip=true,width=0.8\columnwidth]{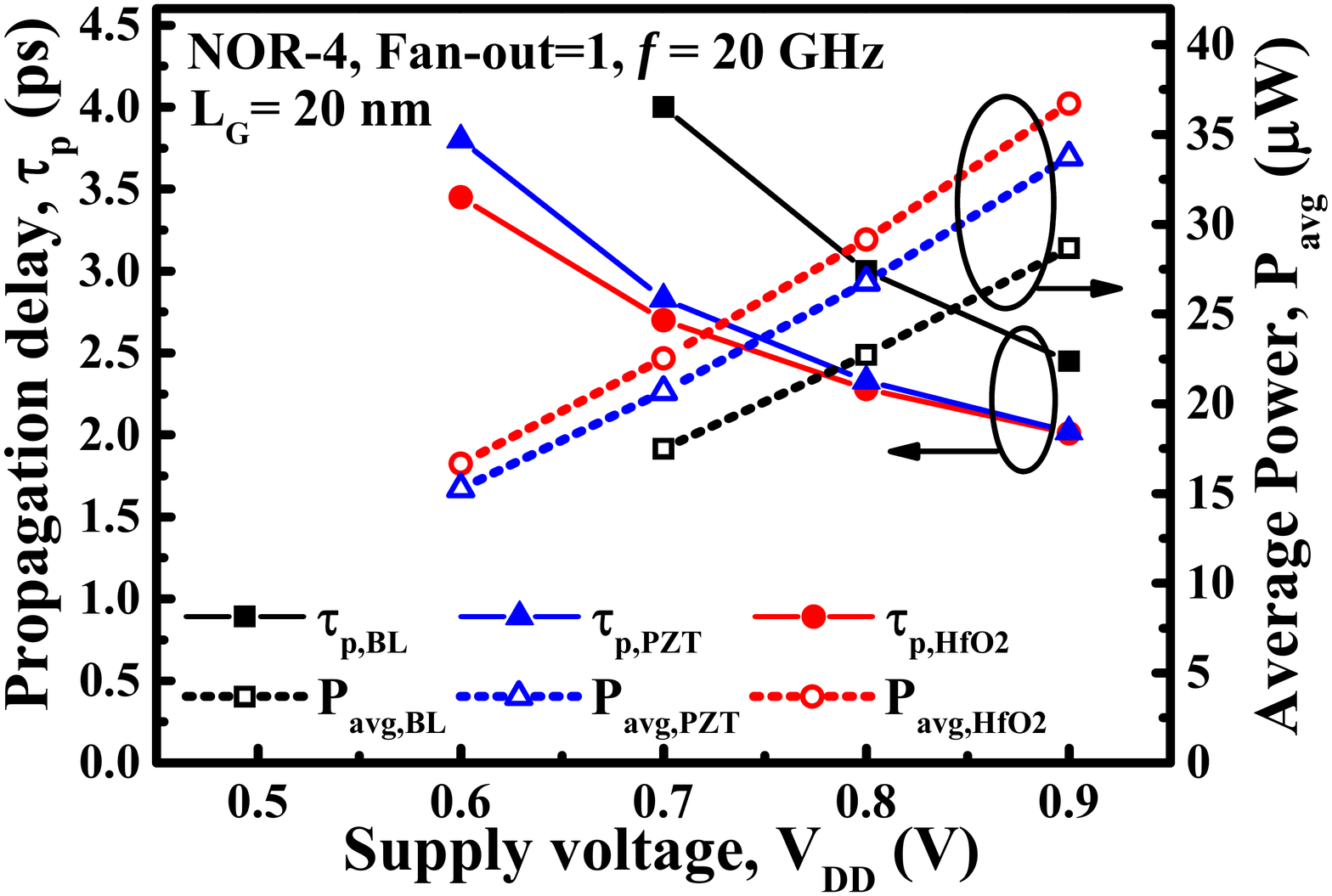}}
\caption{Propagation delay and average power consumption comparison at different V$_{DD}$ for HfO$_{2}$ FDSOI NCFET based NOR-4 gate, PZT FDSOI NCFET based NOR-4 gate and FDSOI MOSFET based NOR-4 gate. The FDSOI NCFET based NOR-4 gates failed to function at 20 GHz at a V$_{DD}$ of 0.5 V. The baseline FDSOI MOSFET based NOR-4 gate failed to function at 20 GHz V$_{DD}$ of 0.5 V and 0.6 V. Fan-out is 1, $\mid$V$_{BS}$$\mid$=1 V.}
\label{nor4_graph}
\end{figure}

The effect of increased fan-in and fan-out of logic gates on performance and average power consumption was also studied. Due to increased current driving capability of FDSOI NCFET devices, the performance of FDSOI NCFET based gates was found to be better with increased fan-in and fan-out. The fan-in was increased from 2 to 4 and 8 with a fan-out of 1. When fan-in was increased to 4 for NAND gates, at a V$_{DD}$ of 0.5 V, HfO$_{2}$ and PZT FDSOI NCFET based NAND gates could be operated till 20 GHz. But, the FDSOI MOSFET based NAND-4 gate could not be operated beyond 5 GHz at the same operating voltage. Fig. \ref{nand4_graph} shows the comparison of performance and average power consumption for NAND-4 gates at different V$_{DD}$ values. Further, HfO$_{2}$ FDSOI NCFET based NAND gates continued to have superior performance when fan-in was increased to 8. While at a V$_{DD}$ of 0.5 V, the baseline FDSOI MOSFET based NAND-8 gates could not be operated beyond 1 GHz, the HfO$_{2}$ and PZT FDSOI NCFET based NAND-8 gates could operate till 10 GHz. Similar results were obtained when fan-in of NOR gates was increased to 4 and 8. The results for NOR-4 gates are shown in Fig. \ref{nor4_graph}.

\begin{figure}[!ht]
\setlength{\abovecaptionskip}{-3pt}
\centerline{\includegraphics[trim=2cm 1cm 0.5cm 2cm clip=true,width=0.8\columnwidth]{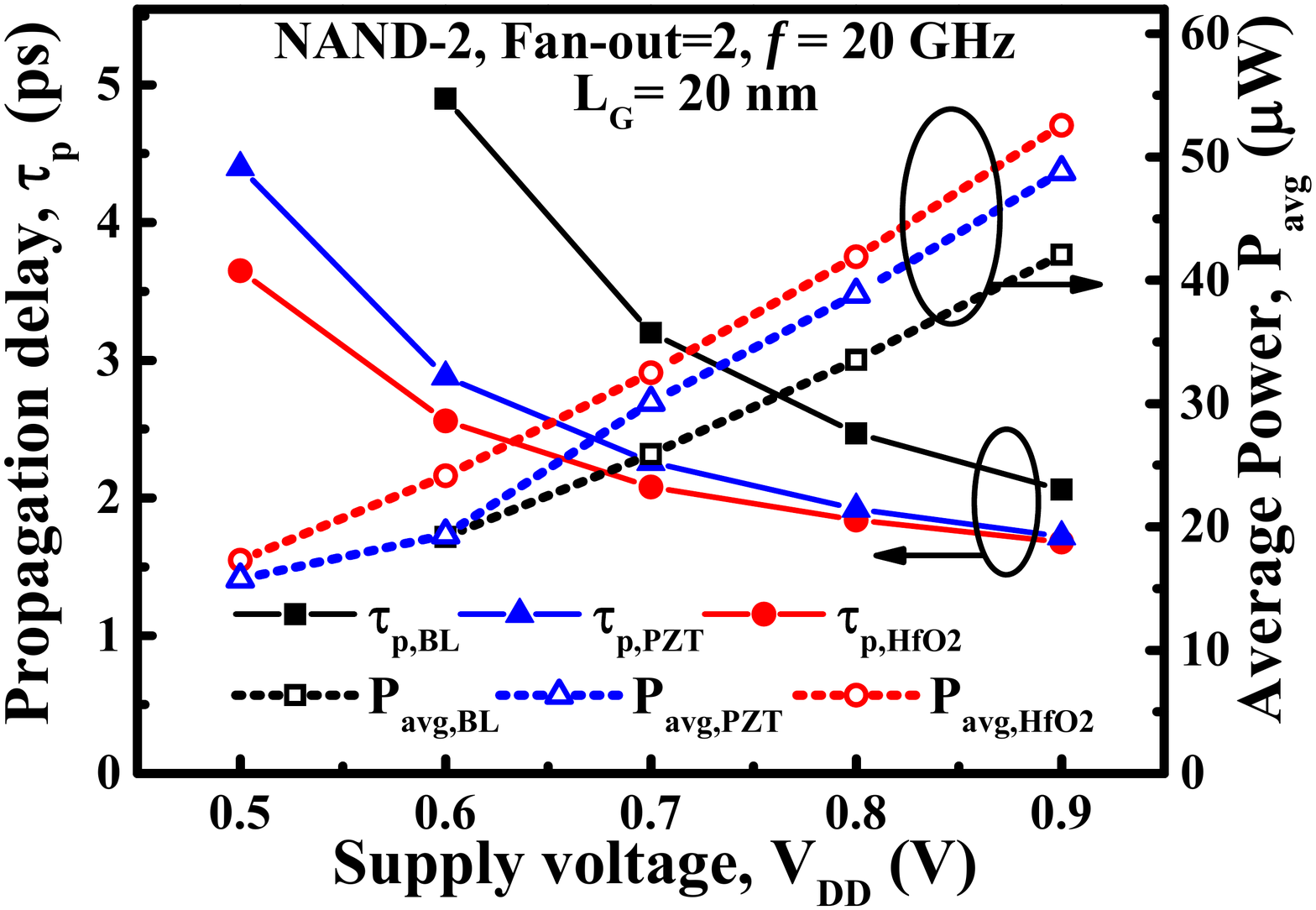}}
\caption{Propagation delay and average power consumption comparison at different V$_{DD}$ for HfO$_{2}$ FDSOI NCFET based NAND-2 gate, PZT FDSOI NCFET based NAND-2 gate and FDSOI MOSFET based NAND-2 gate, each with fan-out of 2. The baseline FDSOI MOSFET based NAND-2 gate with fan-out of 2 failed to function at 20 GHz at a V$_{DD}$ of 0.5 V. $\mid$V$_{BS}$$\mid$=1 V.}
\label{nand2_fan-out=2}
\end{figure}

\begin{figure}[!h]
\setlength{\abovecaptionskip}{-3pt}
\centerline{\includegraphics[trim=2cm 1cm 0.5cm 2cm clip=true,width=0.8\columnwidth]{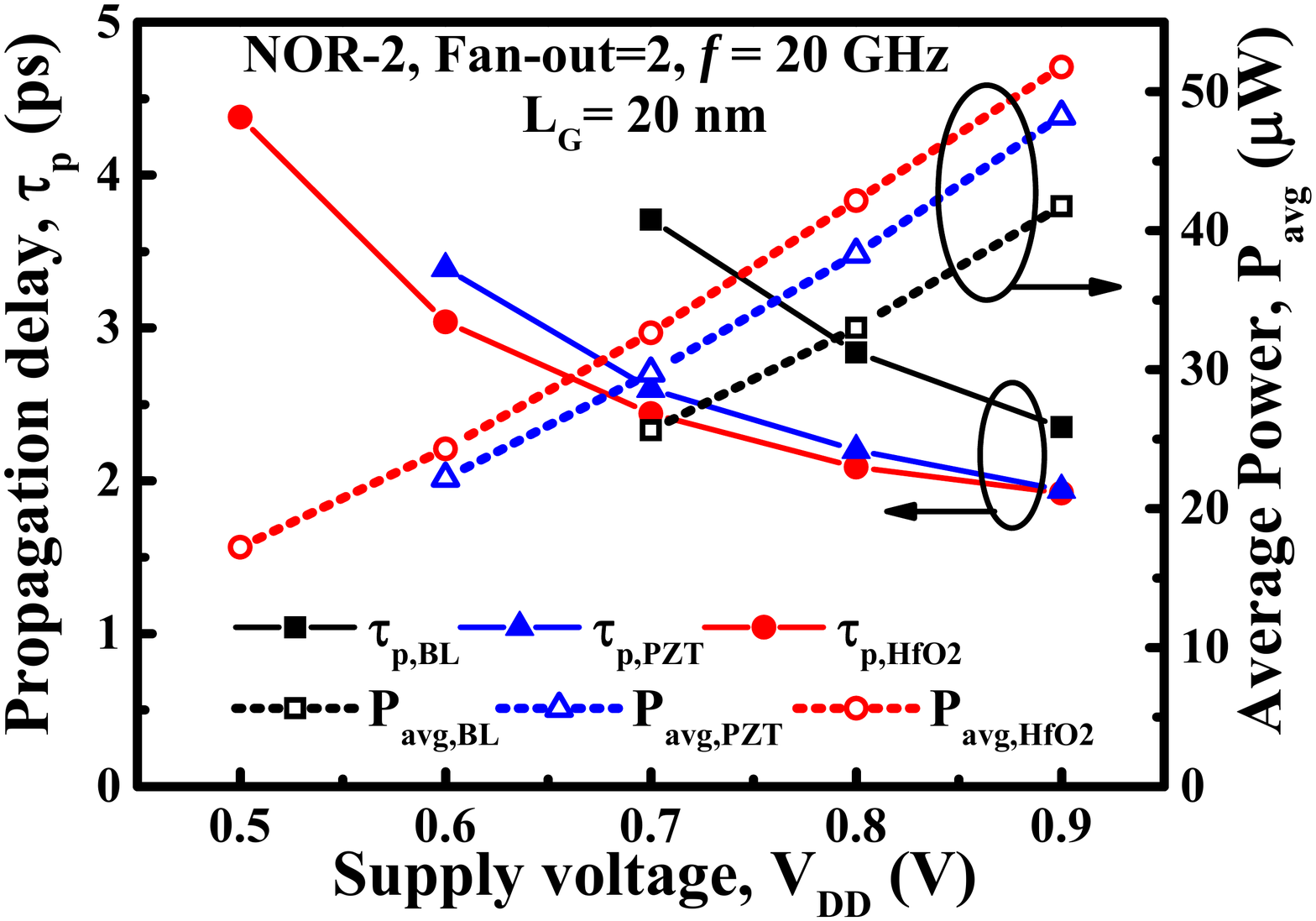}}
\caption{Propagation delay and average power consumption comparison at different V$_{DD}$ for HfO$_{2}$ FDSOI NCFET based NOR-2 gate, PZT FDSOI NCFET based NOR-2 gate and FDSOI MOSFET based NOR-2 gate, each with fan-out of 2. The baseline FDSOI MOSFET based NOR-2 gate with fan-out of 2 failed to function at 20 GHz at V$_{DD}$ of 0.5 V and 0.6 V. The PZT FDSOI NCFET based NOR-2 gate with fan-out of 2 failed to function at 20 GHz at V$_{DD}$ of 0.5 V. $\mid$V$_{BS}$$\mid$=1 V.}
\label{nor2_fan-out=2}
\end{figure}

\begin{table}[ht]
\centering
\setlength{\tabcolsep}{3pt}
\caption{Propagation delay and average power consumption for NAND-2 gates using HfO$_{2}$ FDSOI NCFETs for different T$_{FE}$. The signal frequency is 20 GHz, $\mid$V$_{BS}$$\mid$=1 V.}
\label{nand2_hfo2_table}
\begin{tabular}{|c|c|c|c|c|c|c|c|c|}
\hline
\textbf{V$_{DD}$} & \textbf{$\tau_{p}$} &\textbf{$\tau_{p}$} & \textbf{$\tau_{p}$} & \textbf{$\tau_{p}$} & \textbf{P$_{avg}$} & \textbf{P$_{avg}$} & \textbf{P$_{avg}$} & \textbf{P$_{avg}$}\\
& (BL) & (HfO$_{2}$) & (HfO$_{2}$) & (HfO$_{2}$) & (BL) & (HfO$_{2}$) & (HfO$_{2}$) & (HfO$_{2}$)\\
 & & (5 nm) & (10 nm) & (15 nm) & & (5 nm) & (10 nm) & (15 nm)\\
\textbf{(V)}      & \textbf{(ps)}              & \textbf{(ps)} & \textbf{(ps)} & \textbf{(ps)}  & \textbf{($\mu$W)}     & \textbf{($\mu$W)} & \textbf{($\mu$W)} & \textbf{($\mu$W)}\\
\hline
0.5 & 7.05 & 3.85 & 2.45 & 1.8 & 8.86 & 10.24 & 11.74 & 13.86\\
\hline
0.6 & 3.3 & 2.31 & 1.75 & \textcolor{red}{1.46} & 13 & 14.63 & 16.82 & \textcolor{red}{19.69}\\
\hline
0.7 & 2.15 & 1.75 & \textcolor{red}{1.46} & 1.27 & 17.69 &  19.82 & \textcolor{red}{22.65} & 26.56\\
\hline
0.8 & 1.67 & \textcolor{red}{1.44} & 1.28 & 1.18 & 23 & \textcolor{red}{25.72} & 29.32 & 34.38\\
\hline
0.9 & \textcolor{red}{1.41} & 1.27 & 1.17 & 1.13 & \textcolor{red}{29} & 32.1 & 37.06 & 40.1\\
\hline
\end{tabular}
\end{table}

The fan-out of NAND-2 and NOR-2 gates was increased to 2 and performance and average power consumption of the gates were analyzed as shown in Figs. \ref{nand2_fan-out=2} and \ref{nor2_fan-out=2}. While at a V$_{DD}$ of 0.5 V, HfO$_{2}$ and PZT FDSOI NCFET based NAND-2 gates driving 2 inverters could be operated till 20 GHz, the baseline FDSOI MOSFET based NAND-2 gate driving 2 CMOS inverters could not be operated beyond 10 GHz. Similarly, when NOR-2 gates driving 2 inverters using baseline FDSOI MOSFETs were operated at 0.5 V of V$_{DD}$, they could not function beyond 5 GHz. On the other hand PZT FDSOI NCFET based NOR-2 gates with fan-out of 2 could not operate beyond 10 GHz while HfO$_{2}$ FDSOI NCFET based NOR-2 gates with fan-out of 2 could be operated till 20 GHz at a V$_{DD}$ of 0.5 V. 

\subsection{Effect of change in T$_{FE}$}
The influence of change in T$_{FE}$ on improvement in delay and reduction in average power can be instructive from device design and device fabrication point of view. Table \ref{nand2_hfo2_table} shows the performance of NAND-2 gates with fan-out of 1 for different T$_{FE}$ of HfO$_{2}$. As T$_{FE}$ is increased, the delay improves for a given V$_{DD}$ but the average power consumption also increases. For the same delay of $\sim$1.4 ps, the average power consumption in the baseline FDSOI MOSFET based gate is $\sim$29 $\mu$W. It is reduced by $\sim$11\% at T$_{FE}$ of 5 nm and by $\sim$32\% at T$_{FE}$ of 15 nm. Similar results were obtained for NOR-2 gates also.
%

\section{Comparison of HfO$_{2}$ NCFETs and PZT NCFETs}
As discussed in the preceding sections, HfO$_{2}$ FDSOI NCFET based circuits show better performance than PZT FDSOI NCFET based circuits. This is despite the fact that for FDSOI NCFETs used in our study, T$_{FE}$ of PZT ($\sim$20 nm) was greater than T$_{FE}$ of HfO$_{2}$ ($\sim$10 nm). For NAND-2 gates with fan-out of 1, HfO$_{2}$ FDSOI NCFET based NAND-2 gates are 18\% faster than PZT FDSOI NCFET based NAND-2 gates at the same V$_{DD}$ of 0.5 V. Similar results were obtained for NOR-2 gates also. 

\section{Conclusion}
The paper shows the significance of thin ($\sim$10 nm) HfO$_{2}$ and ($\sim$20 nm) PZT as ferroelectrics in the gate stack of FDSOI NCFETs for high performance, low V$_{DD}$ low-power digital circuits at 20 nm gate length. The study of HfO$_{2}$ FDSOI NCFET based gates and PZT FDSOI NCFET based gates shows that HfO$_{2}$ as a ferroelectric is more promising for high performance, low V$_{DD}$ low-power digital circuits. Further, significant improvement is achieved in the power-delay product by using HfO$_{2}$ FDSOI NCFETs for logic gates. The performance of HfO$_{2}$ FDSOI NCFET based gates for increased fan-in and fan-out was also found to be superior to PZT FDSOI NCFET based gates and baseline FDSOI MOSFET based gates. However, the study did not consider the damping effect of ferroelectrics.

\bibliographystyle{IEEEtran}
\bibliography{References1}

\end{document}